\documentclass{vgtc}                          % final (conference style)
\ifpdf%                                % if we use pdflatex
  \pdfoutput=1\relax                   % create PDFs from pdfLaTeX
  \usepackage{graphicx}                % allow us to embed graphics files
  \DeclareGraphicsExtensions{.pdf,.png,.jpg,.jpeg} % for pdflatex we expect .pdf, .png, or .jpg files
\else%                                 % else we use pure latex
  \usepackage{graphicx}                % allow us to embed graphics files
  \DeclareGraphicsExtensions{.eps}     % for pure latex we expect eps files
\fi%

\graphicspath{{figures/}{pictures/}{images/}{./}} % where to search for the images

\usepackage{microtype}                 % use micro-typography (slightly more compact, better to read)
\PassOptionsToPackage{warn}{textcomp}  % to address font issues with \textrightarrow
\usepackage{textcomp}                  % use better special symbols
\usepackage{mathptmx}                  % use matching math font
\usepackage{times}                     % we use Times as the main font
\usepackage{cite}                      % needed to automatically sort the references
\usepackage{tabu}                      % only used for the table example
\usepackage{booktabs}                  % only used for the table example
\usepackage{comment}
\usepackage{amsmath}                    % added by Kaustav
\usepackage[dvipsnames]{xcolor}         % added by Kaustav
\usepackage{adjustbox}
\usepackage{balance}
\definecolor{mygray}{gray}{0.95}

\newcommand\tab[1][1cm]{\hspace*{#1}}

\newcommand*{\Comb}[2]{{}^{#1}C_{#2}}%

\onlineid{3415}

\vgtccategory{Research}

\vgtcinsertpkg

\title{PRIVEE: A Visual Analytic Workflow for Proactive \\ Privacy Risk Inspection of Open Data}

\author{Kaustav Bhattacharjee, Akm Islam, Jaideep Vaidya, and Aritra Dasgupta
\thanks{K. Bhattacharjee, A. Islam and A. Dasgupta are with NJIT \protect\\
\tab[0.45cm] Email: \{kb526,azi3,aritra.dasgupta\}@njit.edu \protect\\
\tab[0.45cm] J. Vaidya is with Rutgers University \protect\\
\tab[0.45cm] Email: jsvaidya@business.rutgers.edu
}}

\abstract{
Open data sets that contain personal information are susceptible to adversarial attacks even when anonymized. By performing low-cost joins on multiple datasets with shared attributes, malicious users of open data portals might get access to information that violates individuals' privacy.
However, open data sets are primarily published using a release-and-forget model, whereby data owners and custodians have little to no cognizance of these privacy risks. We address this critical gap by developing a visual analytic solution that enables data defenders to gain awareness about the disclosure risks in local, joinable data neighborhoods. The solution is derived through a design study with data privacy researchers, where we initially play the role of a red team and engage in an ethical data hacking exercise based on privacy attack scenarios. We use this problem and domain characterization to develop a set of visual analytic interventions as a defense mechanism and realize them in PRIVEE, a visual risk inspection workflow that acts as a proactive monitor for data defenders. PRIVEE uses a combination of risk scores and associated interactive visualizations to let data defenders explore vulnerable joins and interpret risks at multiple levels of data granularity. We demonstrate how PRIVEE can help emulate the attack strategies and diagnose disclosure risks through two case studies with data privacy experts.
} % end of abstract

\CCScatlist{
  \CCScatTwelve{Human-centered computing}{Visu\-al\-iza\-tion}{Visualization application domains}{Visual analytics};
}

\begin{document}


\firstsection{Introduction}

\maketitle

%% \section{Introduction} %for journal use above \firstsection{..} instead
Accessibility of open data portals~(e.g., NYC open data~\cite{NYCOpenData:online}) is like a double-edged sword. On the one hand, they make institutions and organizations accountable by providing public access to proprietary information. On the flip side, inadvertent data leaks could compromise the privacy of data subjects. Recent research has shown how the lack of checks and balances in the conventional release-and-forget model~\cite{rocher2019estimating} makes it surprisingly easy to breach privacy. An underlying reason for such a high privacy risk is the joinability of multiple open data sets that contain information about people. However, data owners and custodians~(hereafter referred to as defenders) lack effective ways in which joinability risks can be summarized and communicated at the time of data set release or whenever a vulnerability is detected online.

Several recent examples of privacy breach scenarios emphasize the urgent need to address this problem. The Australian Department of Health released \textit{de-identified} medical records for 2.9 million patients (10\% of the population), but researchers were able to re-identify the patients and their doctors using other open demographic information~\cite{culnane2017health}. Passengers' private information might be disclosed through the public transportation open data released by the city municipal of Riga, Latvia~\cite{lavrenovs2016privacy}. Researchers were also able to re-identify the details for 91\% of all the taxis in NYC using an anonymized open taxi dataset and an external dataset~\cite{douriez2016anonymizing}.

Complete automation of the risk evaluation process is not feasible due to several reasons, like the presence of noisy metadata and the requirement for human expertise. Noisy metadata hinders the automatic profiling of these datasets. The various definitions and temporal nature of privacy risks, owing to the intermittent release of new datasets, point to the necessity for a human-in-the-loop approach, where defenders can configure and update risk computation techniques based on evolving compliance needs. 

To address this critical need, we conducted a design study with urban informatics and data privacy researchers to develop a \textbf{proactive risk inspector} that is privy to the sensitive information that can be leaked before and after dataset release in urban, open data portals. PRIVEE, the visual analytic workflow resulting from this design study process, acts as a data-driven risk confidante and informer for the defender in the analysis loop. PRIVEE emulates potential attack scenarios and enables defenders to triage risky dataset combinations and ultimately diagnose the severity of disclosed information through dataset joins. A defender can thus proactively check for risks while releasing a dataset or depend on PRIVEE to be alerted when new vulnerabilities emerge owing to newly available, joinable data. 

As the first contribution of this design study, we characterize the problem of disclosure evaluation and develop a set of visual analytic tasks that can be executed in a workflow to detect, calibrate, and inform data defenders about disclosure risks~(Sections \ref{section:red_teaming}, \ref{section:tasks}). These tasks, developed in collaboration with privacy experts, emerged when we analyzed the problem through the lens of an adversary and developed several attack scenarios. We observed that it is possible to breach the privacy of open datasets using these scenarios, thus corroborating the findings of NYC taxi data in a larger scope where we can find information about data subjects~\cite{douriez2016anonymizing}.
As our second contribution, we designed the visualizations required for implementing the PRIVEE workflow and let defenders explore and interpret risks at the \textbf{metadata level}, triage vulnerable dataset groups and corresponding high-risk \textbf{joinable dataset pairs}, and ultimately reason about the severity of the information disclosed at a \textbf{record-level}~(Section~\ref{section:privee_design_overview}). The design of these techniques is rooted in the idea of automation with transparent explanations which are responsive to user-controlled risk configurations~(Sections \ref{section:joinable_groups},~\ref{section:joinability_risk},~\ref{section:disclosure_detection}). Finally, we present an interactive interface to help data defenders execute the workflow and demonstrate its effectiveness in the end-to-end diagnosis of disclosure~(Sections~\ref{section:case_study_vulnerability}, \ref{section:case_study_id_disclosure}) through two case studies with domain experts. 
\section{Background \& Related Work} \label{section:related_work}

% \aritra{It is not a good practice to write a long, undivided section on Related Work. Also, purely going by the size of Related Work, it should be easily doubled}
% \kaustav{Updated this section to add some background and merged it with the related work}

\begin{figure}
\begin{center}
\includegraphics[width=0.8\columnwidth]{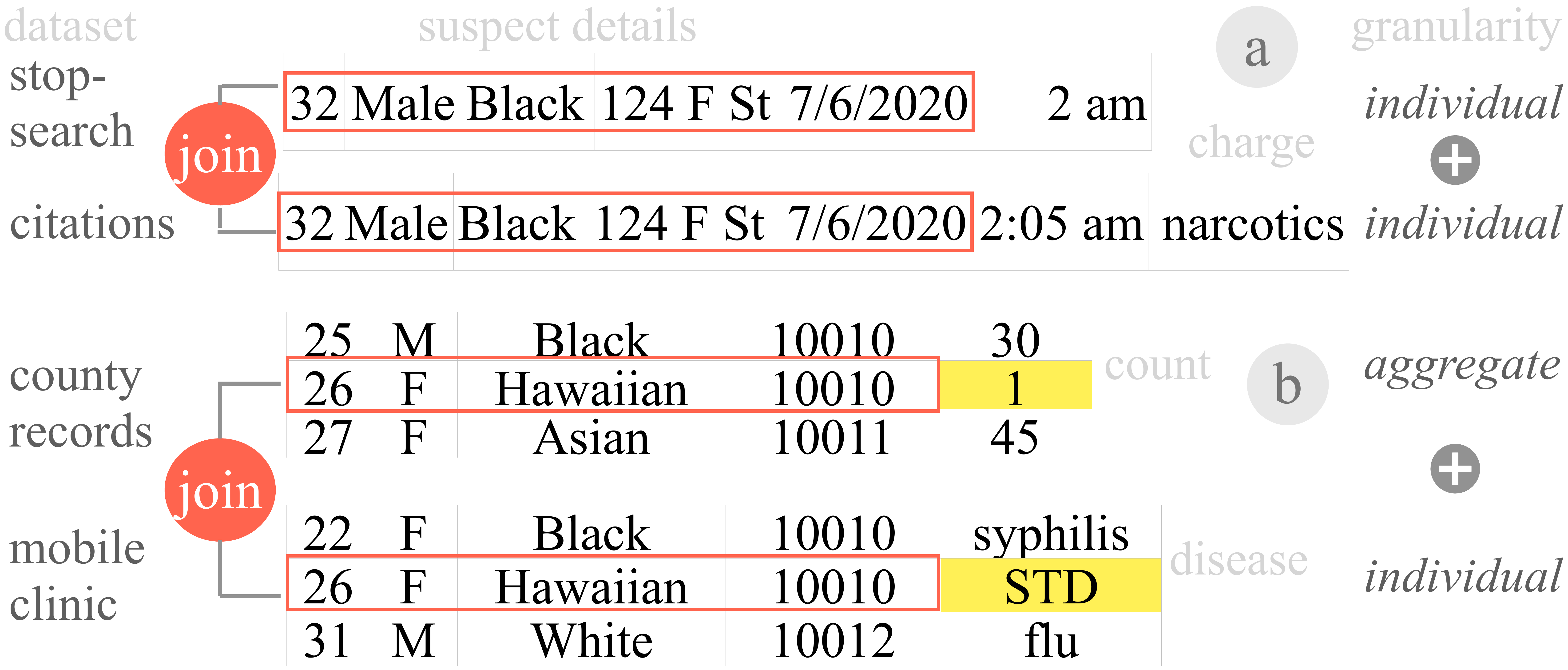}
\vspace{-3.5mm}
 \caption{\label{figs:disclosure_example}\textbf{Open Data Disclosure risks using real datasets:} (a) Two \textit{individual-level} de-identified open datasets can be joined using quasi-identifiers like age, sex, race, location, and date in order to identify an individual and reveal sensitive information about them like their citation charge. (b) \textit{Aggregated-level} datasets can cause disclosure risk when a record with a meager value can be joined with another individual-level dataset, e.g., the only 26-year-old Hawaiian female living in a particular zip code has STD.}
\end{center}
\vspace{-8mm}
\end{figure}

The Open Data Charter was signed by the leaders of the G8 nations in 2013, leading to the increasing adoption of datasets that can be freely used, re-used, and redistributed by anyone, commonly referred to as \textit{open datasets}~\cite{opendatadefinition, Ourhisto94:online}. Though these are generally anonymized before release, joining two anonymized datasets using protected attributes can lead to the disclosure of sensitive information. In this context, \textit{direct identifiers} are those protected attributes that can directly link to and identify an individual from a dataset (like name, id, SSN), while \textit{quasi-identifiers} are those protected attributes that individually do not uniquely identify an individual but when combined with others, can identify an individual~(e.g., age, race, gender, location). 
%Quasi-identifier like the age of a person cannot point to a specific individual, but a combination of other quasi-identifiers like race, gender, location, job title, etc. can disclose the information about the individual since the presence of an individual of a certain age, gender, race in a particular location can be unique enough to identify the individual. 
Disclosure risks can be mainly of two types: \textit{identity disclosure}, where the data consumer knows who the individuals are, and \textit{attribute disclosure}, where the values of different quasi-identifiers or sensitive attributes (like disease, salary, etc.) are revealed. Figure~\ref{figs:disclosure_example} shows two examples of identity and attribute disclosures using open datasets related to traffic stop-search, police citation, mobile clinic, and county health records. %Figure~\ref{figs:disclosure_example}a  shows an example where records from two different de-identified open datasets, like traffic stop-search  and police datasets, can be linked using quasi-identifiers to identify an individual and reveal sensitive information like their citation charge (possession of narcotics). 
%The quasi-identifiers age, sex, race, and location are used to link the individual from both the datasets. 
%Another example (Figure ~\ref{figs:disclosure_example}b) shows that the low presence of a certain group of people in a open aggregated record-level dataset can also lead to the disclosure of sensitive information like disease when joined other individual-level open datasets based on quasi-identifiers like age, gender, race and zip code.

A suite of anonymization methods~\cite{fung2010privacy} exists to address the problem of linking among public and private datasets, for example, between Census data and hospital records. The most promising among those methods is the notion of differential privacy~\cite{dwork2014algorithmic} that the US Census has recently adopted~\cite{dwork2019differential, ruggles2019differential}. However, besides US Census data, which is just one of the sources of openly available data about people and their behavior, there are now a plethora of open data portals. As mentioned earlier, the adoption of open data is based on the promise of transparency and utility, as depicted by the FAIR principles~\cite{wilkinson2016fair}, and at the same time, on the need for adherence to emerging privacy laws~\cite{nouwens2020dark}.
%The adoption of open data is a recent phenomenon, with the promise of transparency and utility as depicted by the FAIR principles~\cite{wilkinson2016fair}, and at the same time, with the need for adherence to emerging privacy laws~\cite{nouwens2020dark}.
The unrestricted availability of open data~\cite{green2017open} naturally raises the question: what if datasets within the open data ecosystem are linked even without other sensitive information from private datasets? 
%Researchers have recently expressed the need to investigate privacy issues in open data portals. %Lavrenovs and Podins showed how the privacy of passengers can be violated through the public transportation open data, released by the city municipal of Riga, Latvia\cite{lavrenovs2016privacy}. 
Recent studies have demonstrated how even heavily anonymized datasets can be used to re-identify about $99\%$ of Americans~\cite{rocher2019estimating}. Re-identification or the disclosure of sensitive information is a challenge that has been previously explored by multiple researchers~\cite{de2015unique, de2013unique, zang2011anonymization}. 

\begin{comment}
The risk of re-identification can be considered a temporal function dependent on the datasets available at the time of analysis~\cite{sekara2021temporal}. 
%Sekara et al. have recently performed a study that shows that this risk varies over time for the re-identification using a smartphone app usage dataset with publicly available information\cite{sekara2021temporal}. 
However, data owners often practice the "release-and-forget" model where datasets, once released, are not analyzed further for potential privacy risks concerning the newly released datasets~\cite{ohm2009broken, rubinstein2016anonymization}. 
This may lead to an inadvertent breach of privacy
%and leak of sensitive information 
when new datasets are joined with the existing ones.
%This may lead to an inadvertent breach of privacy when new datasets are released, which can leak sensitive information when joined with the existing datasets. 
Thus, it is necessary for the data owners to proactively monitor the risks of their datasets concerning other datasets. %available through the various open data portals.
%and keep checking regularly.
\end{comment}
Data owners often practice the \textbf{release-and-forget model} where datasets, once released, are not analyzed further for potential privacy risks concerning the newly released datasets~\cite{ohm2009broken, rubinstein2016anonymization}. 
However, the risk of re-identification can be considered a temporal function~\cite{sekara2021temporal}, thus requiring proactive monitoring of the risks. We developed the PRIVEE workflow and visualization interface with the specific goal of realizing a defender-in-the-loop analytical framework that can be privy to the disclosure risks or possibility of accidental leakage of sensitive information whenever new datasets are released. Though the target users of PRIVEE are mainly data custodians or data owners, even data subjects~\cite{bhattacharjee2020privacy} can use the workflow to inspect how vulnerable their identity or personal information might be in the presence of multiple, linked data sets. 
We use the concept of dataset joinability~\cite{miller2018making} in the presence of quasi-identifiers as a means for calibrating disclosure risks that are communicated using interactive visualizations throughout the PRIVEE workflow. 
Commercial tools like Google Cloud Data Loss Prevention (DLP) also help visualize the disclosure risk of a particular dataset using the quasi-identifiers~\cite{GoogleDLP:online}. 
While we do find other examples of visualization techniques for expressing disclosure risks of individual datasets~\cite{kao2017using,dasgupta2019} and sensitive information~\cite{kum2019enhancing,dasgupta2014opportunities}, \textit{interactively visualizing disclosure risk among joinable open datasets} is essentially an open problem that we address in PRIVEE.

\section{Problem Characterization} \label{section:red_teaming}

% \aritra{I think we need an example like in our \textbf{CGF} paper with tables to demonstrate joinability risk.}
% \kaustav{Added an example in the background and relevant work section}

\begin{figure*}[t]
\begin{center}
\includegraphics[width=0.9\textwidth]{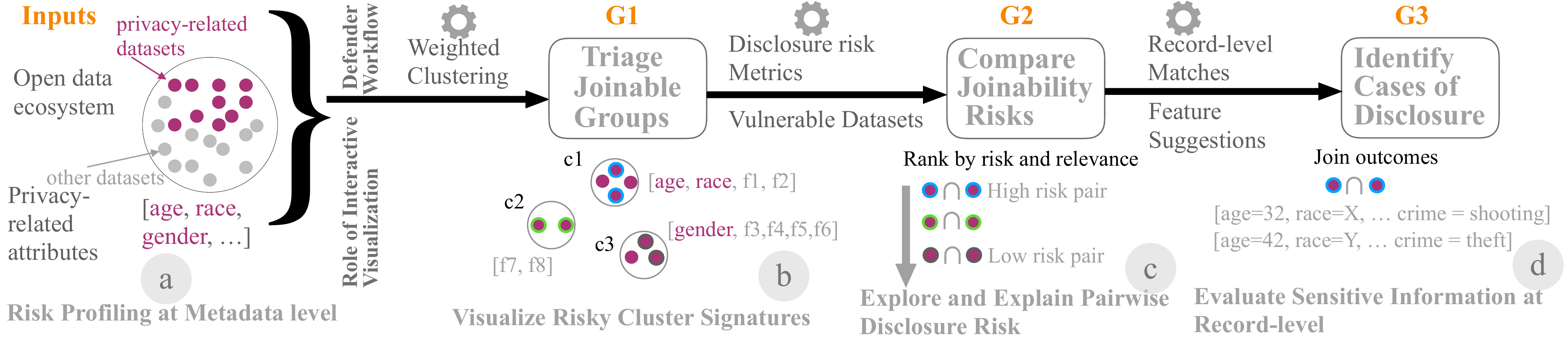}
\vspace{-4mm}
 \caption{\label{figs:PRIVEE}\textbf{PRIVEE is an end-to-end risk inspection workflow for open datasets} that informs the defender in the analytical loop about potential disclosure risks in the presence of joinable datasets. Interactive visualization plays a crucial role in bootstrapping the risk inspection process via risk profiling, triaging and explaining risk signatures, and ultimately detecting instances of true disclosure at a record level. Colored borders track datasets across the goals.}
\end{center}
\vspace{-8mm}
\end{figure*}
To understand the requirements for addressing the disclosure risks through the linking of open datasets, we decided to conduct a \textit{red-team exercise} with the help of researchers in data privacy and urban informatics. A red-team exercise can be generally defined as a structured process to better understand the capabilities and vulnerabilities of a system by viewing the problems through the lenses of an adversary~\cite{zenko2015red}. 
% In the context of security and privacy, red-team exercises usually follow the cyber kill chain by playing the role of an ethical hacker and emulating the possible attack scenarios~\cite{hutchins2011intelligence}.
%With the help of researchers in data privacy and urban informatics, we performed a red-teaming exercise by inspecting the open datasets for vulnerabilities. 
We engaged in a cold-start exploration process, followed by a more focused exploitation of datasets with privacy-related attributes, to develop a shared mental model of the problems related to the vulnerabilities and understand the functional requirements of a system addressing these vulnerabilities. We used the data sketches method and shared ideas about the different strategies with our collaborators~\cite{lloyd2011human} and explored multiple attack scenarios.
%The intuition here was that the datasets with universally known quasi-identifiers, like age, race, gender, etc., can lead to the disclosure of sensitive information when joined with other such datasets. 

%\par \noindent \textbf{Exploiting vulnerable entry points:} 
% \par \noindent \textbf{Attack Scenario exploiting vulnerable entry points:} 
Red-team exercises generally follow the cyber kill chain, which starts with the initial reconnaissance step, where attackers try to \textit{find vulnerable entry points} into any target system~\cite{hutchins2011intelligence}. Following this step, we bootstrapped our red-teaming activity by first defining an initial set of privacy-related attributes, like age, race, gender, 
%age group, 
and location, to name a few. During our initial exploration, we collected  $39,507$ datasets from around $500$ data portals and observed through an automated analysis that about $5404$ datasets have some combinations of quasi-identifiers. We filtered out datasets related to non-human objects, leading to the retrieval of a seed set of $426$ datasets, including $151$ individual record-level~(e.g., records of people committing crimes) and $275$ aggregated record-level~(e.g., college records) datasets~\cite{PRIVEE_NJIT_Dataset}. 
Analysis of these datasets led to interesting observations where some of the datasets have a highly skewed distribution of records across different categories of the quasi-identifiers. For example, the dataset \textit{Whole Person Care Demographics 2}~\cite{WholePersonDemographics2:online} from the \textit{County of San Mateo Datahub portal}\cite{SMCDatahub:online} has only one record for a 26-year-old Hawaiian female, similar to the example shown in Figure~\ref{figs:disclosure_example}b. This can lead to identity disclosure and may leak sensitive information when joined with other datasets.% individual record-level datasets. 
%Similar examples were also observed in the datasets from other portals.
% , like a dataset from the \textit{Austin Public Health} open data portal~\cite{austinpublichealth:online} has only one record for a 16-year old Hispanic female who also uses Medicare insurance. 

% \aritra{Too much overlap between RQs and Gs- why are both needed? I did not find any need for RQs to be honest. We can instead have design requirements derived for each task.}
% \kaustav{Removed the RQs and introduced design requirements in the introduction section for each of the core sections (5,6,7,8)}

% This led to the formation of the first requirement:
% \mybox{\textbf{RQ1: Find vulnerable entry points through metadata}\\
% (a) Use metadata to filter through the datasets
% (b) Find datasets which have records vulnerable  for disclosure}

%\par \noindent \textbf{Exploiting dataset joins}:
% \par \noindent \textbf{Attack Scenario exploiting dataset joins:} 
After building an initial collection of vulnerable datasets, we aimed to understand the consequence of an attacker joining them and accessing sensitive information. In this context, we would like to highlight that join is a fundamental operation that connects two or more datasets, and joinability is the measure to determine if two datasets are linkable by any number of join keys~\cite{dong2021efficient,chia2019khyperloglog}. When these \textit{join keys coincide with protected attributes} like age, race, location, etc., the outcome of the join can potentially reveal sensitive information about an individual or even disclose the individual's identity.
As a next step in the red-teaming exercise, we randomly selected vulnerable pairs of datasets from multiple open data portals~\cite{NYCOpenData:online, KansasCityOpenData:online, DallasOpenData:online}
% , like NYC Open Data~\cite{NYCOpenData:online}, Open Data Kansas City~\cite{KansasCityOpenData:online}, City of Dallas Open Data~\cite{DallasOpenData:online}, etc. 
and analyzed them for~\textit{joinability risks}, in terms of what kind of sensitive information may be leaked by these joins. 
% Another requirement was identified here:
% \mybox{\textbf{RQ2: Find and rank joinable datasets}\\
% (a) Find which datasets can be joined among the candidate datasets (b) Identify high-risk joins}

%These two datasets had $39$ shared attributes along with quasi-identifiers like age, race, and sex. 

Several iterations of the selection of joinable pairs and join keys % along with different combinations
%of the shared attributes as 
led to the discovery of disclosure between the datasets \textit{Juvenile Arrests} and \textit{Adult Arrests} from the \textit{Fort Lauderdale Police Open Data Portal}\cite{FortLauderdaleOpenData:online}. We observed that two individuals, aged $15$ and $21$, mentioned separately in these datasets, were involved in the same incident of larceny on $20$\textsuperscript{th} March $2018$, at the Coral Ridge Country Club Estate, Fort Lauderdale, similar to the example in Figure~\ref{figs:disclosure_example}a. We repeated this exercise and found other examples where dataset joins ultimately led to disclosures.% either identity disclosure or disclosure of sensitive information.

\section{Visual Analytic Goals and Tasks} \label{section:tasks}
% \aritra{We need a PRIVEE interface figure to complement this section}
% \kaustav{Added the interface figure}
\begin{figure*}[t]
\begin{center}
\includegraphics[width=\textwidth]{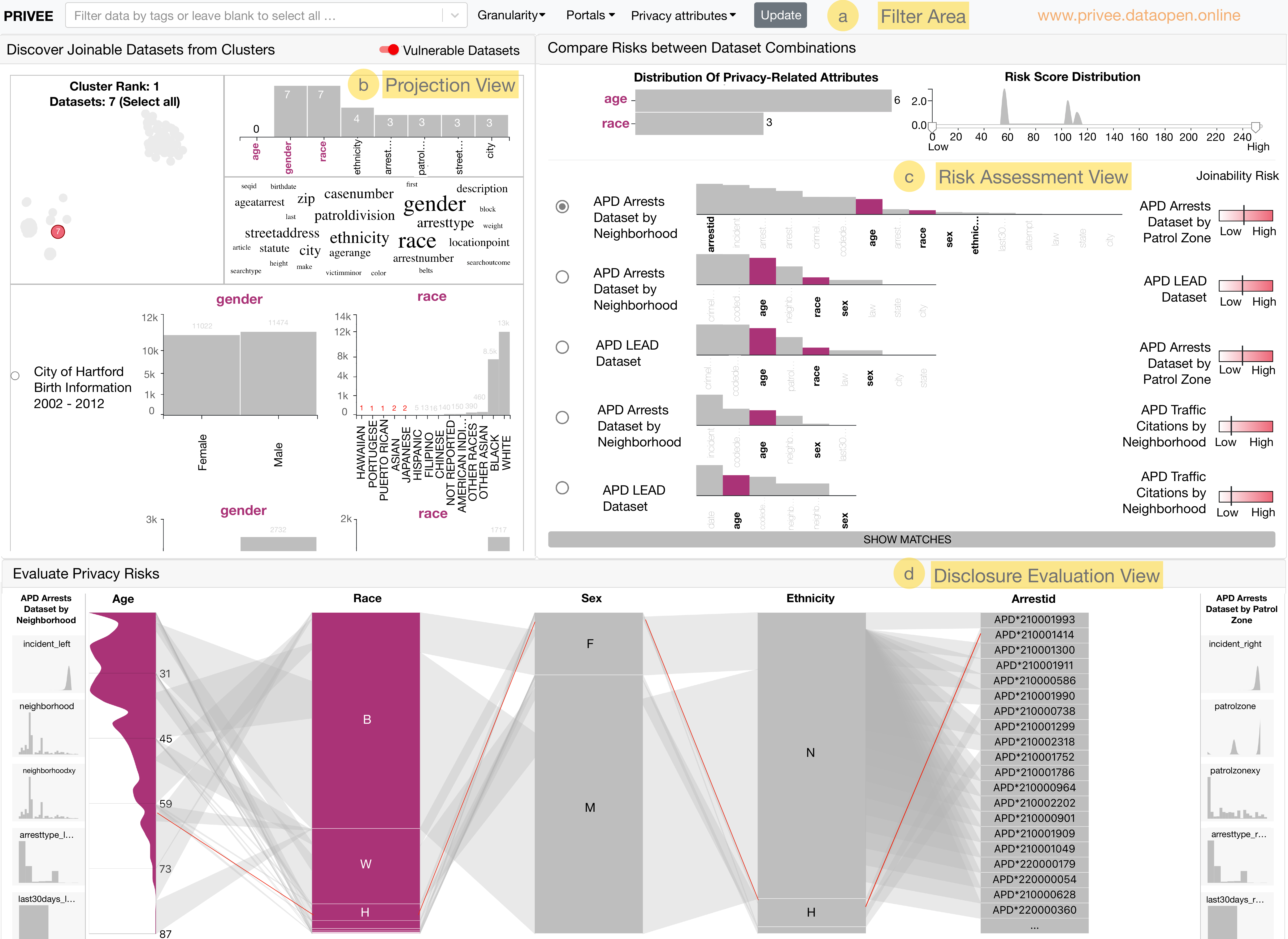}
\vspace{-7mm}
 \caption{\label{figs:PRIVEE_Overview}\textbf{Interface Design:} The design of PRIVEE comprises rich interaction among filters and multiple views: (a) Filter area helps select datasets based on metadata like tags, data granularity, and privacy-related attributes; (b) Projection View lets the defenders compare the signatures of different joinable groups of datasets and evaluate vulnerable data distributions; (c) Risk View helps compare the risk for dataset pairs and select the high-risk pairs; (d) Disclosure Evaluation View helps to analyze the matching records for potential disclosures. }
\end{center}
\vspace{-9mm}
\end{figure*}
The results from the red-teaming exercise confirmed our intuition that datasets with quasi-identifiers, when linked together, can potentially divulge sensitive information. Analyzing the functional requirements, we, together with our collaborators, concluded that totally automating the risk evaluation process is infeasible as human intervention is necessary at multiple stages of risk definition, interpretation, and subsequent exploration of the dataset combinations at high risk. To formulate a solution, we collaboratively developed PRIVEE, a visual risk inspection workflow in which defenders can proactively engage to stay one step ahead of the attackers~(Figure \ref{figs:PRIVEE}).
%for these anticipated scenarios~(Figure \ref{figs:PRIVEE}).

PRIVEE is motivated by protecting the most vulnerable data sets against data join attacks. 
%We designed PRIVEE to emulate the attack scenarios on behalf of the data defenders by mapping the functional requirements to visual analytic goals and tasks. 
%Discussions with our collaborators in data privacy and urban informatics led to the eventual development of the visual analytic workflow. 
The workflow serves the dual purpose of: i) observing the open datasets to detect potential privacy vulnerabilities and ii) being a trusted informer for the data defenders that can visually explain and communicate disclosure risks while encouraging a deeper exploration of the attack and defense strategies. 
Automating the analysis of the disclosures directly at the record level can be an alternative, but this may lead to a seemingly infinite number of combinations to explore.
Our streamlined workflow, developed from the experience gained during this design study process, will help the data defenders focus on a set of highly vulnerable datasets, thus reducing the number of combinations to be explored.
In this section, we first describe the inputs and then define the high-level goals of the PRIVEE workflow in order to map them to the corresponding visual analytic tasks ultimately realized in a web-based interface.

\par \noindent \textbf{{Inputs to the workflow:}} We initiate our defense strategy on the seed set of privacy-related datasets, which are about people as the data subjects, that we collected during the red teaming activity. While collecting these datasets, we followed the universally accepted common quasi-identifiers like age, race, gender, etc., with the notion that an open data ecosystem should, at a minimum, protect against attacks using these well-known quasi-identifiers. 
%Suppose we are to imagine an open data ecosystem that provides a minimum privacy guarantee. In that case, that ecosystem should, at a minimum, protect against attacks using these well-known quasi-identifiers. 
%We note that this is not an exhaustive collection of privacy-related datasets.

\begin{comment}
After carefully curating the metadata from the seed datasets, we observed that there is no standard nomenclature for the attributes across the different data portals. This lack of standardization established the importance of creating a metadata dictionary, starting with the well-known set of quasi-identifiers like age, race, gender, and location, and focusing on other similar attributes while providing defenders the guidance and flexibility to define other privacy-related attributes.
%This set of  privacy-related attributes
These attributes and the datasets selected based on their metadata serve as the inputs to the PRIVEE workflow~(Figure \ref{figs:PRIVEE}a).
\end{comment}
After carefully curating the metadata from the seed datasets, we observed that there is no standard nomenclature for the attributes across the different data portals. This lack of standardization established the importance of creating a metadata dictionary,
%starting with the well-known set of quasi-identifiers like age, race, gender, and location, and 
focusing on the well-known quasi-identifiers while providing defenders the guidance and flexibility to define other privacy-related attributes.
%This set of  privacy-related attributes
These attributes and the datasets selected based on their metadata serve as the inputs to the PRIVEE workflow~(Figure \ref{figs:PRIVEE}a).

\par \noindent \textbf{{G1: Triage Joinable Groups:}} Candidate datasets for inspection selected from the initial input can be of the order of tens or hundreds. Finding all possible combinations of dataset joins among them is computationally expensive.
Moreover, the large set of join outcomes will not lend well to human interpretation of risk. Also, during the red-teaming exercise, we observed that the risky datasets could also be construed from the datasets with vulnerable data distributions.
Therefore, the next tasks in the defender's workflow are to focus on groups of datasets that can be joined and then triage those groups based on risk indicators:
%, leveraging the following task:

\textbf{T1}: \textit{Explore cluster signatures:} As shown in Figure \ref{figs:PRIVEE}b, this task lets defenders explore cluster signatures in terms of presence~(clusters c1, c3) or absence~(cluster c2) of the privacy-related attributes and their overall semantics. Involving the defender ensures that their inputs influence the algorithms used for grouping, using weighted clustering. They can thus control the triaging process by judging the groups' risks and privacy relevance. This task ultimately helps them select clusters of interest for further inspection of joinability risks.
%, thus fulfilling the functional requirement \textit{\textbf{RQ2a}}, which is about finding joinable groups of datasets.

\textbf{T2}: \textit{Find vulnerable datasets based on data distributions:} The red-teaming exercise highlighted the presence of disclosure risk in datasets with a highly skewed records distribution across different categories of the quasi-identifiers. This task helps to distinguish between the most vulnerable and other datasets by inspecting a high likelihood of finding unique records for given quasi-identifiers.

\par \noindent \textbf{{G2: Compare Joinability Risks:}}  Once a cluster of datasets is prioritized for inspection as part of G1, defenders would like to compare joinable pairs of datasets in this group that may potentially disclose sensitive information. To achieve this goal, we use disclosure risk metrics to automatically suggest risky pairs based on their feature profiles and then visualize those suggestions so defenders can interpret the metrics. 
% But during the red-teaming exercise, we observed that the risky pairs could also be construed from the datasets with vulnerable data distributions. 
The following task achieves this:

% \textbf{T2}: \textit{Find vulnerable dataset based on data distributions:} The red-teaming exercise highlighted the presence of disclosure risk in datasets with a highly skewed records distribution across different categories of the quasi-identifiers. This task allows defenders to distinguish between the most vulnerable and other datasets by inspecting a high likelihood of finding unique records for a given set of quasi-identifiers. 

\textbf{T3}: \textit{Explore and Explain Disclosure Risks:} This task focuses on pairs of datasets that can be ranked using multiple disclosure risk metrics. Within those rankings, we want to use visual cues that directly explain: which features are responsible for high risk, the differences between high and low-risk pairs, and if other features should augment the defender's definition of privacy relevance.
%This satisfies the requirement \textbf{RQ2b}, which mandates identifying high-risk joins.

%An automatic risk scoring of the dataset pairs is an ideal solution for this case, but may not be sufficient since  
% PRIVEE also augments the selection of the joining attributes by helping users navigate through a sea of attributes common to both the datasets in a dataset combination.
%Several datasets can be joinable through a varying extent, thus arises the need for analyzing the  risks among these datasets.
%\aritra{add description}

%\kaustav{Use in workflow}will help a defender save time and effort which may otherwise be wasted if they have to analyze every possible pair of datasets. For example, if 10 datasets are selected, $\Comb{10}{2}$, i.e., $45$ dataset combinations need to be analyzed. We propose a ranking algorithm based on the risk score of the dataset combinations along with the entropy of their common attributes.

\par \noindent \textbf{G3: Identify cases of disclosure:} Once dataset pairs are selected as part of G2, defenders would like to understand the severity of the join outcomes.
%, as per the requirement \textit{\textbf{RQ3}}.
Fully automating this process may lead to many scenarios where the disclosures are less concerning and do not warrant any significant change in the defense strategies. To provide more control to defenders in their diagnosis of cases of actual disclosure, the tasks required to accomplish this goal are:

\textbf{T4}: \textit{Detect matching records across data sets:}
Matching records are the records present in both datasets in a pair.
%Matching records are records from each of the datasets of a pair of datasets and can be joined using the values from a set of shared attributes. 
The main objective of this task is to detect lower frequencies of matching records, which may lead to the disclosure of sensitive information about an individual or disclose their identity. 

\textbf{T5}: \textit{Augmenting the risky feature set with suggestions:}
One way of discovering disclosures is finding attributes that have the same values for all the records of the joined datasets. For example, joining two hospital datasets may reveal that all the patients common in both the hospitals are treated for cancer, leading to attribute disclosure for these patients.
%, thus rendering this dataset join unsafe
In this task, we suggest a set of attributes that may be highly related to the joining attributes, thus helping the users augment the feature set for the dataset join.

\section{Design Overview}\label{section:privee_design_overview}

The design of PRIVEE is motivated by the need for a transparent explanation and evaluation of the risk inspection process. We implemented a web-based interface that enables data defenders to iterate between multiple entry points, evaluate the reasons for the dataset joinability and analyze disclosure risks for different combinations of datasets and attributes. In this section, we provide an overview of the design requirements for realizing the aforementioned visual analytic goals and tasks.
\textbf{An interactive version of the interface} for PRIVEE may be accessed through the Chrome browser at \url{http://privee.dataopen.online/}. 

\noindent \textbf{\textit{Risk Profiling at metadata level}}: PRIVEE helps to analyze the datasets' risk profiles through a filter bar, located conveniently at the top of the interface~(Figure~\ref{figs:PRIVEE_Overview}a), which contains a search option for the different tags and options to select the data portals and the dataset granularity. During the initial page load, this filter bar is positioned at the center of the page in order to avoid overwhelming the user with the search results. Defenders can select any combination of the tags from the tags search option, which is enriched with a modified bar chart showing the frequency distribution of the tags. Though the tags are sorted in descending order, the grey bar in the background (achieved by tweaking a linear-gradient bar) provides an idea of the frequency distribution of these tags among all the collected datasets. Privacy-related attributes can also be selected using filters. 
% These attributes are colored violet (plum kingdom), implementing a color scheme maintained throughout all the views of PRIVEE.

%PRIVEE also allows the defenders to select a combination of the privacy-related attributes and update the dataset selection accordingly. The position of the filter bar enhances the multi-view interface so that the data defenders can iteratively select any combination of datasets for their analysis. Thus, PRIVEE helps in the guided selection of datasets based on their risk profile by leveraging their metadata \textbf{(T1)}.
\noindent \textbf{\textit{Triaging joinable groups}}: In order to fulfill G1, PRIVEE employs a set of visualizations to help the data defenders triage the joinable groups from the datasets selected using their metadata. This includes a projection plot, a word cloud, and a bar chart depicting the attributes' frequency, as illustrated in Figure~\ref{figs:PRIVEE_Overview}b. This combination of visualizations is repeated for the different groups of joinable datasets. Though PRIVEE automates the grouping of the datasets, these visualizations provide the data defender a transparent method to understand the group signatures and update the groups based on their domain knowledge and definition of privacy relevance. 

% Design choices for this \textit{Projection View} have been discussed in Section~\ref{section:joinable_groups}. Defenders can select any group of datasets from this view for further inspection.% in the Risk Assessment View.

\noindent \textbf{\textit{Finding vulnerable datasets}}: PRIVEE helps the data defenders select vulnerable datasets by showing a distribution of the values of the privacy-related attributes through a combination of histograms (for numerical attributes) and bar charts (for categorical attributes), as shown in Figure~\ref{figs:PRIVEE_Overview}b. This combination is repeated for each dataset, ranked according to their degree of vulnerability. It is also responsive to the privacy-related attributes selected through the filter area. The vulnerable categories for these attributes and their labels are shown in bright red to help defenders efficiently select vulnerable datasets.

\noindent \textbf{\textit{Comparing Joinability Risk}}: PRIVEE automatically computes the possible pairs from the datasets selected from either Projection View or the Vulnerable Datasets View and ranks them according to their joinability risk. The visual cues, shown in Figure~\ref{figs:PRIVEE_Overview}c, help the data defender compare different datasets and select the high-risk pairs on a priority basis. Overall information about the risk score distribution allows flexible selection of dataset pairs of varying risk. 

% The design choices and the metrics calculation for this \textit{Risk Assessment View} have been further discussed in Section~\ref{section:joinability_risk}. Dataset pairs, and a selection of attributes for the join key serves as the input for evaluating and identifying disclosures.

\noindent \textbf{\textit{Identifying disclosures}}: The disclosure of sensitive information can depend on multiple factors, subject to evaluation by the data defender. In this \textit{Disclosure Evaluation View}, as shown in Figure~\ref{figs:PRIVEE_Overview}d, PRIVEE lets the data defender analyze the matching records generated for a specific dataset pair and a join key selected from the Risk Assessment View. PRIVEE also suggests other features to help the defenders select a better join key, helping them understand the relationship between different attributes and possible disclosures.

\section{Triage Joinable Groups (G1)}\label{section:joinable_groups}
%The degree of joinability between the candidate datasets may vary based on multiple factors. Hence, 
Data defenders need to analyze the degree of joinability between datasets. Hence, the design requirements for addressing tasks T1 and T2 are to develop human-in-the-loop clustering methods responsive to multiple definitions of privacy relevance, along with transparency in analyzing cluster signatures. This enables defenders to develop a mental model of the context and the degree of the potential vulnerability of subsequent joins. In this section, we discuss the analytical methods and visualizations to find and triage the joinable groups.

\subsection{Weighted clustering for finding joinable datasets}

\noindent \textbf{\textit{Converting Data Attributes to Word Embeddings}}:
The joinability of two datasets is a function of \textit{shared attributes}. Hence, the datasets with similar attributes should be more joinable. Attribute names in open datasets are often noisy and inconsistent, making it computationally difficult to perform a binary search for the presence or absence of certain attributes.  We focus on the idea that similar attribute names can capture the semantic similarity among multiple datasets that might have a similar context. We use a word-embedding approach that simultaneously satisfies the need to capture datasets' joinability and their semantic similarity. \textit{Word embeddings} can be defined as real-valued, fixed-length, dense, and distributed representations that can capture the lexical semantics of words~\cite{bakarov2018survey, almeida2019word}. Hence, we converted the data attributes into their corresponding word embedding form using Python's spaCy library~\cite{Spacy} and created a vector representation for the attribute space of each dataset. The vectors with a smaller distance between themselves signify datasets with similar attributes, hence more joinable.

%These vector word embedding representations of the attributes help to perform multiple useful operations like addition, subtraction, distance calculation, etc.; hence we used them to form a vector representation for the attribute space of each dataset~\cite{almeida2019word}. All the datasets were represented by a corresponding vector representation of their attributes at this stage. The vectors with a smaller distance between themselves signify datasets with similar attributes, hence more joinable.

\noindent \textbf{\textit{Adding Weights for Privacy-related attributes}}:
At this stage, all the data attributes have equal importance in the vector representation of a dataset; hence, datasets with attributes like \textit{version}, \textit{version number}, etc. may be marked similar to each other. However, these attributes may not have much significance in the context of privacy. Hence, we decided to add weights to some of the privacy-related attributes identified from the seed dataset corpus. Attributes like \textit{age}, \textit{race}, \textit{gender} and \textit{age at arrest} were selected, and adding more weights to these attributes signifies that datasets having these attributes may be marked as more joinable. Any disclosure using these datasets can be considered a high risk, which will help further triage the datasets.%pairs.

Cosine similarity is widely used to measure the similarity between words and documents~\cite{thongtan2019sentiment, dai2015document}. However, word embeddings are mere representations of the words, and multiplying them with numeric weights would not increase the cosine similarity between two datasets. Hence we introduced a \textit{weight vector} where we assign a weight if the privacy-related attributes selected by the data defender are present in the dataset. If a data defender selects the privacy-related attributes [\textit{age, gender, race}], then the corresponding weight vector for a dataset with only the age and gender attributes would be $[x,0,x]$, where $x$ represents the weight assigned to the privacy-related attributes. We concatenate these weight vectors with the corresponding word embedding vectors to get the final vector representation of each dataset. 
%More details about the experiments conducted to identify the privacy-related attributes and the optimal weights can be found in the supplementary materials.

\noindent \textbf{\textit{Projecting the datasets and finding Clusters}}:
Each dataset is now represented by a vector with more than $300$ elements/dimensions, and comparing these datasets using a 2-D or 3-D plot would be challenging if all the dimensions were used. Hence we used the t-SNE dimensionality reduction algorithm to reduce these into two-dimensional vectors ~\cite{van2008visualizing}. %which will be easy to plot on a 2-D scatterplot~\cite{van2008visualizing}. 
A 2-D projection of the datasets might not readily reveal dataset groupings. Hence, we experimented with clustering algorithms like KMeans~\cite{vassilvitskii2006k}, DBSCAN~\cite{ester1996density,schubert2017dbscan}, Birch~\cite{zhang1996birch}, and OPTICS~\cite{ankerst1999optics, schubert2018improving}. After a careful analysis of the clusters' quality and the cluster density scores, we selected the DBSCAN algorithm.
%for PRIVEE. 
%Additional details about the experiment can be found in the supplementary materials.

\noindent \textbf{\textit{Evaluating the clusters}}:
There can be multiple groups of similar/joinable datasets, which would lead to the creation of multiple clusters. A data defender may find it challenging to evaluate all of these clusters. Hence we have employed a few cluster evaluation techniques to triage these clusters \textbf{(T1)}. 

One of such metrics is the \textit{Calinski-Harabasz Index} which is defined as the ratio of the between-cluster dispersion and the inter-cluster dispersion,  where dispersion means the sum squared distance between the samples and the barycenter~\cite{calinski1974dendrite}. A higher score signifies that the different clusters are far away, implying better cluster formation. We designed an experiment to evaluate the difference in the results from this metric along with other metrics like Silhouette Score~\cite{rousseeuw1987silhouettes} and Davies-Bouldin Index~\cite{davies1979cluster} and selected the Calinski-Harabasz Index since we observed that it could efficiently guide defenders in finding meaningful, joinable datasets. Further details about this experiment can be found in the supplementary materials.
%We perform automated calibration of cluster quality using the Calinksi-Harabasz Index metric that ensures we can efficiently guide defenders in finding meaningful, joinable datasets.

\noindent \textbf{\textit{Finding vulnerable data distributions}}:
A particular cluster can have multiple datasets with vulnerable data distributions, leading to the disclosure of sensitive information when joined with other individual record-level datasets. Hence, we found such data distributions and ranked these datasets according to their degree of vulnerability~\textbf{(T2)}.

In order to evaluate the degree of vulnerability, we first analyzed all the datasets and created the record points for the privacy-related attributes present in them. Record points are the unique categories for a specific attribute, while \textit{vulnerable record points} are those record points that have very few records for them, as shown in Table \ref{table:record_points}. These datasets are then sorted based on the number of such vulnerable record points present and the frequency of the most vulnerable record point. The intuition here is that a dataset with more vulnerable record points is more prone to disclosure risk using these privacy-related attributes.

\begin{table}[h]%[h!]
    \centering
    \begin{tabular}{|c|c|}
    \hline
     \textbf{Record points} & \textbf{Description} \\ 
     \hline
     [``age”, 11, 1] & For age=11, there is only 1 record \\ 
     \hline
     [``age”, 15, 5] & For age=15, there are 5 records \\
     \hline
     [``gender”, ``F”, 2] & For gender=``F", there are 2 records \\
    \hline
    \end{tabular} %\break
    \caption{Sample record points}
    \label{table:record_points}
\end{table}
\vspace{-4mm}

\subsection{Visualizing joinable group signatures}
\begin{figure}
\begin{center}
\includegraphics[width=\columnwidth]{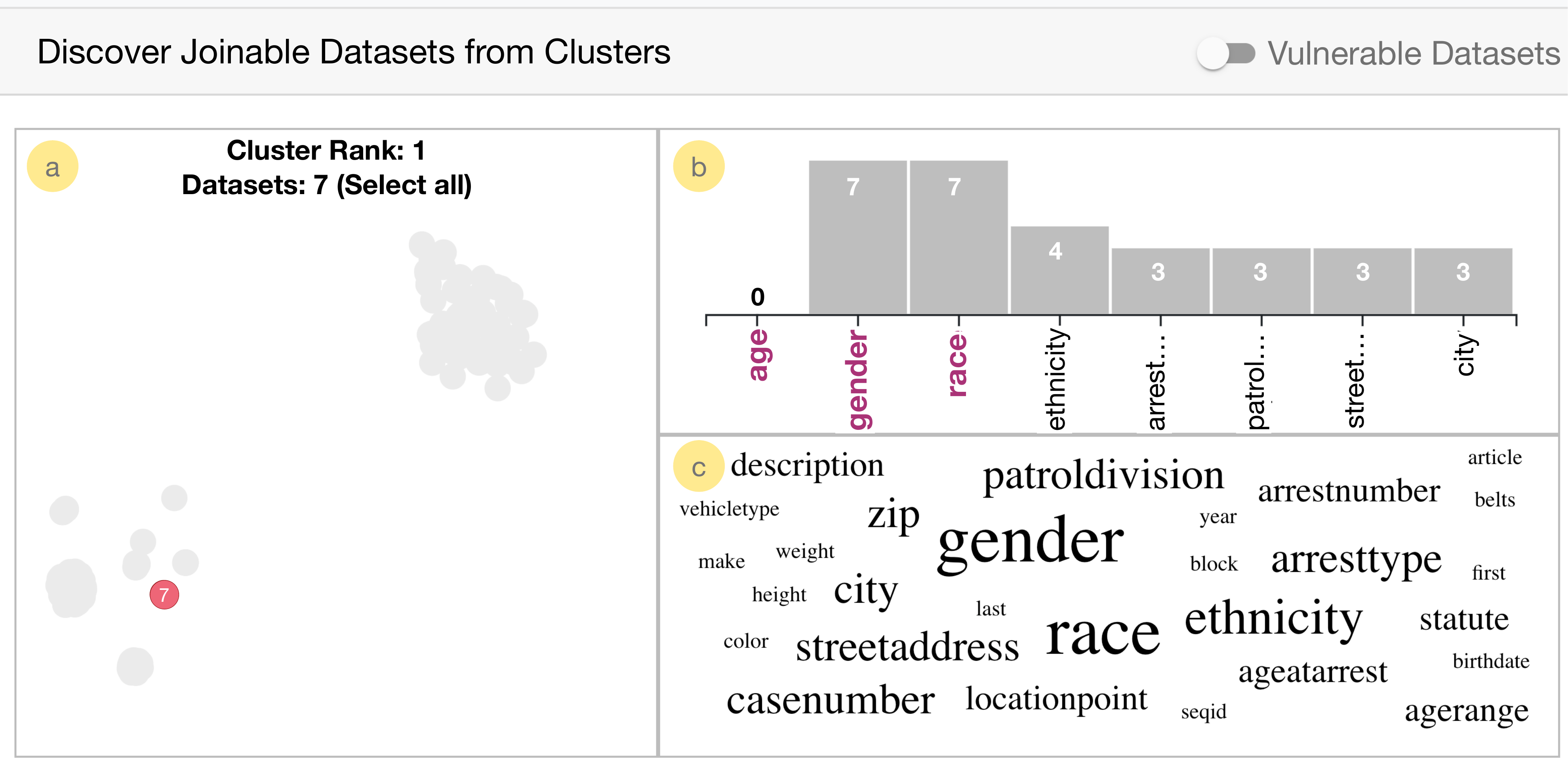}
\vspace{-6mm}
 \caption{\label{figs:projection_view}\textbf{Projection View:} A group of joinable datasets is represented using (a) a projection plot. The (b) frequency distribution bar chart and (c) word cloud for the attributes of a group of joinable datasets help in the transparent explanation of the group signatures.}
\end{center}
\vspace{-10mm}
\end{figure}

%The design of PRIVEE is inspired by \textit{Shneiderman's Visual Information-Seeking Mantra}: Overview first, zoom and filter, then details-on-demand~\cite{shneiderman2003eyes}.
We designed the Projection View to provide an overview of the datasets and the joinable groups~\textbf{(T1)} and perform an automatic evaluation of the vulnerable data distributions of the datasets in each joinable group~\textbf{(T2)}. Data defenders can review the group signatures through the different components of the Projection View and update the parameters to see the details and the data distribution of the datasets that match their mental model of privacy relevance. The components of these views are described as follows:

\noindent \textbf{\textit{Joinable groups}}:
Given a set of datasets selected based on their metadata, 
% and vulnerable data distributions, 
defenders need to find groups of datasets that can be joined together. The analytical process is performed automatically by PRIVEE, leading to the formation of joinable clusters, which are represented using a multi-dimensional projection plot, as illustrated in Figure~\ref{figs:projection_view}a. Here, a red dot represents an individual record-level dataset in a particular cluster, while the grey dots represent the datasets not in that cluster. During this design study, we realized that some of the datasets are highly joinable due to their similarity in the attribute space, which would cause overlapping of the dots in a cluster. Hence, the overlapping datasets are represented by a single dot with the number of overlapping datasets inscribed in it. For example, Figure~\ref{figs:projection_view}a shows a cluster of seven highly similar datasets represented using a red dot.
This view contains multiple projection plots, where each plot represents a group of joinable datasets. It helps the data defender quickly compare the different groups from a single view. The dual color encoding scheme (red-grey) helps visually differentiate between the datasets in a group and the other datasets. Initially, a scatterplot with different colors for the different clusters was also considered for this view. However, it was realized that it is challenging to assign perceptually different colors to each cluster when the number of clusters is large, due to the limits of perception. Hence, a multiple plot design approach was chosen with the two-color encoding scheme.

\noindent \textbf{\textit{Transparent explanation of joinability and vulnerability}}:
Understanding the cluster signatures is crucial in understanding the reason behind the genesis of a joinable group~\textbf{(T1)} and the presence of data vulnerabilities~\textbf{(T2)}. Since we have construed these dataset groups based on the similarity in their attribute space, it is essential to understand the frequency of the attributes present in these groups. Hence, bar charts become the natural choice for displaying the most frequent attributes in a group and their frequency, as illustrated in Figure~\ref{figs:projection_view}b. These bar charts are sorted according to the attribute frequency, yet the frequencies of the privacy-related attributes are shown first. The vulnerable datasets are also represented using bar charts (for categorical attributes) / histograms (for numerical attributes) for each of the privacy-related attributes present in them. However, bar charts can have the limitation of visual scalability where only a certain number of bars can be shown due to space constraints~\cite{eick2002visual}. In order to overcome this limitation, we also introduce word clouds of the attributes, as shown in Figure~\ref{figs:projection_view}c. All the attributes present in at least two datasets in a joinable group are represented in this word cloud, with the size channel representing their frequency. 

%Analyzing the difference in frequency distribution can help the data defenders understand the reason behind the formation of different clusters. 
The bar chart in Figure~\ref{figs:projection_view}b 
%clearly 
explains the similarity of the datasets since all seven of these datasets have \textit{gender} and \textit{race} attributes, thus transparently explaining the group signatures. Besides overcoming the visual scalability limitation of the bar chart, the word cloud also helps the data defenders look for other attributes of interest that may have a lower frequency but have much larger relevance in the context of privacy. For example, attributes like \textit{victim age} and \textit{offender age} may not be significant for a general user; however, a data defender working with law enforcement may find them interesting since these attributes are used in the police datasets. PRIVEE enables the data defender to update the default selection of the privacy-related attributes, which triggers a re-rendering of the whole Projection View, thus automatically calculating new groups of joinable datasets with extra weightage to the newly added privacy-related attributes \textit{victim age} and \textit{offender age}. Together, these Projection View components enable human-in-the-loop dataset grouping that is adaptive to various definitions of privacy relevance by transparently displaying measures to evaluate cluster signatures.

%% -------------- New Section --------------
\section{Compare Joinability Risks (G2)}\label{section:joinability_risk}
Dataset groups from the Projection View can lead to multiple pairwise combinations of datasets, where the data defenders need to analyze each pair for their joinability risk. Hence, the design requirement for addressing G2 is to facilitate efficient visual comparison of the risk profile of dataset pairs and guide defenders towards focusing on high-risk dataset pairs. In this section, we describe the metrics that can help a data defender quantify the risk of joinability between the candidate datasets and the subsequent use of visual cues to compare and prioritize the joinable pairs.
%There can be multiple pairwise combinations of datasets from a group of joinable datasets. Moreover, the vulnerable datasets can also be joined with other individual record-level datasets to find possible disclosures. For example, a group of $10$ datasets can lead to $45$ possible pairwise combinations, and a vulnerable dataset can be linked with all the individual record-level datasets, resulting in $151$ pairwise combinations. In order to help the data defenders triage these combinations, we have implemented three metrics \textbf{(T4)}:
\subsection{Metrics for Joinability risk comparison}\label{sub_section:metrics_joinability_risk}
Multiple metrics that can help the data defenders compare the joinability risks between different dataset pairs were explored during the design study process. In this subsection, we define the mathematical formulas for the different metrics that highlighted the joinability risks better and were selected as part of the PRIVEE workflow. \\
\noindent \textbf{\textit{Metric based on attribute profile}}:
Shannon’s entropy is a measure of the uncertainty of a random variable~\cite{cover1999elements}. It has been widely used as a privacy metric~\cite{serjantov2002towards,diaz2002towards,oganian2003posteriori,alfalayleh2014quantifying}, as higher entropy %of an attribute in a dataset 
signifies more unique values for that attribute, thus resulting in higher disclosure risk. Hence, we used this metric to help defenders find joinable attributes for a pair of datasets. For a pair of datasets (say A and B), we first calculated Shannon's entropy of each of their shared attributes according to equation~\ref{equation:entropy_A} and kept their maximum as the entropy score for that attribute.
%We used equation~\ref{equation:entropy} to calculate the final entropy score for that attribute.
The intuition here is that the attributes with higher entropy can be offered as suggestions to the defender for the join key.
%;hence the shared attributes are sorted according to their final entropy score in descending order.
\begin{equation}
    H(X_{J}) = -\sum_{i=1}^{n} P(x_{J_{i}}) \ln P(x_{J_{i}})
    \label{equation:entropy_A}
\end{equation}

% \begin{equation}
%     H(X_{B}) = -\sum_{j=1}^{m} P(x_{B_{j}}) \ln P(x_{B_{j}})
%     \label{equation:entropy_B}
% \end{equation}
% \begin{equation}
%     \text{entropy score} = \max\{H(X_{A}), H(X_{B})\}
%     \label{equation:entropy}
% \end{equation}
where $X_{J}$ represents attribute X in dataset J ($J \in \{A,B\}$), $H(X_{J})$ represents the entropy of an attribute present in dataset J while $x_{J_{i}}$ represents each category of the attribute $X_{J}$ in dataset J.
% \begin{description}
% \item [$X_{J}$] = attribute X in dataset J, where $J \in \{A,B\}$
% \item [$H(X_{J})$] = entropy of an attribute present in dataset J
% \item [$x_{J_{i}}$] = each category of the attribute $X_{J}$ in dataset J
% \end{description}

\noindent \textbf{\textit{Metric based on dataset pairs in a join}}:
Since the joinability of two datasets depends upon the number of shared features/attributes between them, the joinability risk score can be calculated as a function of the number of shared attributes and the number of privacy-related attributes between a pair of candidate datasets. The formulae for the \textit{joinability risk score} can be defined as follows:
\begin{equation}
    \text{risk} = \alpha * p + (c-p)
    \label{equation:risk_score}
\end{equation}
where $\alpha$ is the empirical risk ratio (a constant), \textit{p} is the number of privacy-related attributes and \textit{c} is the number of shared attributes.

% \begin{description}
% \item [$\alpha$] = a constant, representing the empirical risk ratio
% \item [p] = the number of privacy-related attributes,
% \item [c] = the number of shared attributes
% \end{description}
The joinability risk score depends on the empirical risk ratio, and to determine its value, we designed an experiment to calculate the risk scores of all the possible combinations of joinable pairs from the seed datasets ($\Comb{426}{2} = 90,525$ combinations). 
%This experiment was repeated multiple times, each time with a different value of the risk ratio, ranging from $1$ to $100$. 
We observed that the value $\alpha = 50$ works well to separate the dataset pairs with privacy-related attributes and pairs without them; hence the empirical risk ratio was fixed at the value of $50$. 
%If two datasets have $15$ shared attributes and only $2$ of them are privacy-related attributes, the risk score can be calculated as $113$, as per equation \ref{equation:risk_score}. 
%A higher joinability risk score signifies that this dataset pair is highly joinable, which can be a step closer to finding disclosures.
We have included further details about this experiment in the supplementary materials.

\subsection{Visual risk assessment}
PRIVEE uses multiple visual analytic components to encode the joinability risk metrics, and these components together form the Risk Assessment View. This subsection describes how we map these metrics with the components of this view so that data defenders can pro-actively analyze the risk between the candidate datasets.
%PRIVEE's goal is to compare the joinability risks of different dataset pairs (G3). Defenders can either select a joinable group of datasets or a vulnerable dataset to update the Risk Assessment View. However, the number of possible pairwise combinations of the candidate datasets can be pretty high; hence PRIVEE can help a data defender analyze the risk of these datasets at the dataset and feature levels. The Risk Assessment View uses visual analytic components to encode these metrics so that data defenders can pro-actively analyze the risk between the candidate datasets. This view also implements Shneiderman's mantra since it provides the on-demand details for the datasets chosen from the Projection View or the Vulnerable Datasets View.\\
\begin{figure}[t]
\begin{center}
\includegraphics[width=\columnwidth]{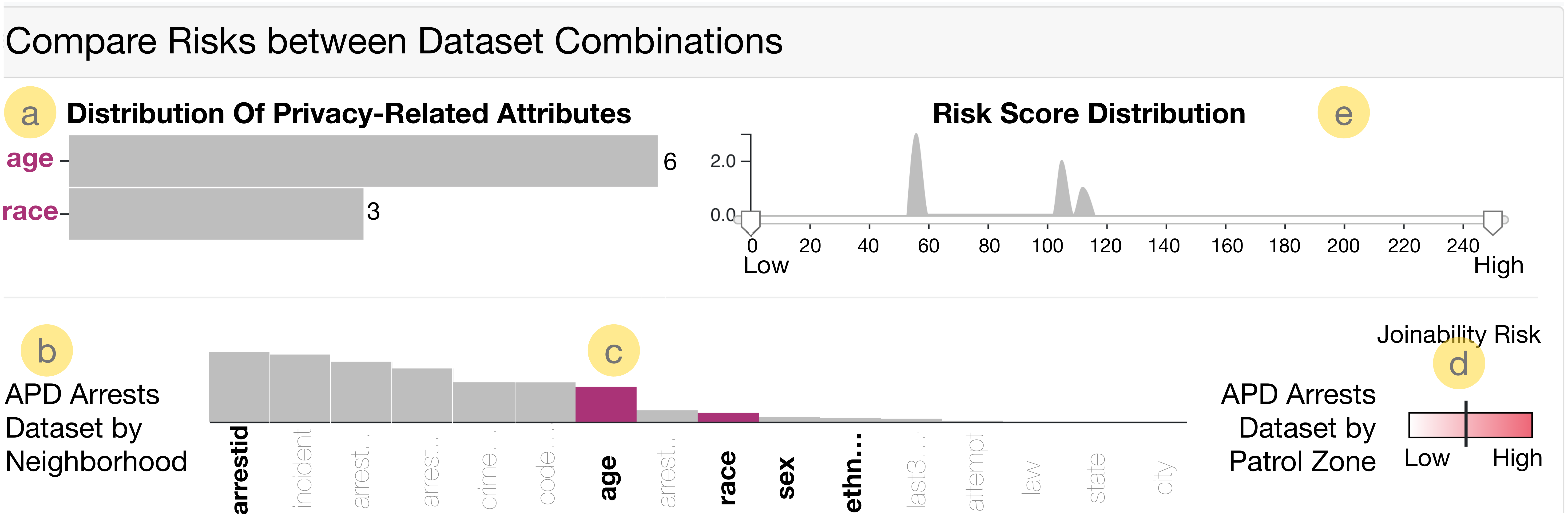}
\vspace{-4mm}
 \caption{\label{figs:risk_view}\textbf{Risk Assessment View}: (a) The distribution of privacy-related attributes can affect the joinability risks between (b) dataset pairs. Data defenders can compare the risk between these pairs by analyzing the (c) sorted bar chart showing the shared attributes and the joinability risk score represented by the (d) risk score bar. They can use the (e) risk score distribution histogram to focus on the dataset pair of their interest.}
\end{center}
\vspace{-9mm}
\end{figure}

\noindent \textbf{\textit{Comparing shared attributes set}}:
The shared attributes' entropy metric encodes the attribute profile information, potentially highlighting if an attribute should be included in the join key. In the Risk Assessment View, these attributes and the entropy are represented using a descending \textit{sorted bar chart} between the dataset names,  as illustrated in Figure~\ref{figs:risk_view}c. The horizontal position shows the different attributes, while the vertical position encodes the entropy of these attributes. The bars for the privacy-related attributes are colored in violet (plum kingdom), while the other bars were colored in grey, thus following the similar colorblind-safe two-color strategy used in the other views. %Other attributes, which are not shared between the dataset pairs, are hidden by design but can be accessed by clicking on the dataset names, if required.
During an initial design iteration, each shared attribute was represented using a small rectangular box, with each box containing the attribute name in it.
%and the boxes were then sorted in the descending order of their entropy. 
However, we realized that this design leads to the loss of information about the difference in entropy between the different shared attributes. This led to the current design of the sorted bar charts where the data defender can analyze the entropy, select any number of the shared attributes as the join key for the dataset pair and evaluate them for disclosures. 
%The selected attributes are colored in golden yellow to facilitate disclosure evaluation using join keys of different risk profiles, thus enabling interactivity between the views in a multi-view interface like PRIVEE.

\noindent \textbf{\textit{Comparing risks}}:
Each dataset pair (Figure~\ref{figs:risk_view}b) is represented with a combination of the following components: dataset names, shared attributes, and the joinability risk bar. 
% These pairs are primarily sorted according to the dataset relevance score, followed by a secondary sorting according to the risk score. 
These pairs are sorted according to the risk score.
Thus, a top-ranked dataset pair would imply higher chances of joinability. In order to highlight the joinability risk score between the dataset pairs, the Risk Assessment View has a \textit{joinability risk bar} for each dataset pair \textbf{(T3)}, as shown in Figure~\ref{figs:risk_view}d. This bar is filled with a linear gradient between the grey and red colors, representing low-risk and high-risk dataset pairs. The exact risk score is highlighted using a black vertical bar. The choice of the colors, following the two-color scheme used across the different views in PRIVEE, helps express the joinability risk score on a scale of low to high scores. % \textbf{(G2)}
This view also shows an overview of the shared privacy-related attributes and the risk score distribution between the dataset pairs using a horizontal bar chart and a histogram (Figure~\ref{figs:risk_view}a and Figure~\ref{figs:risk_view}e). PRIVEE also automatically selects the joining attributes based on their entropy and privacy relevance, which the data defender can further augment.

\section{Identifying disclosures (G3)} \label{section:disclosure_detection}
\begin{figure*}[t]
\begin{center}
\includegraphics[width=\textwidth]{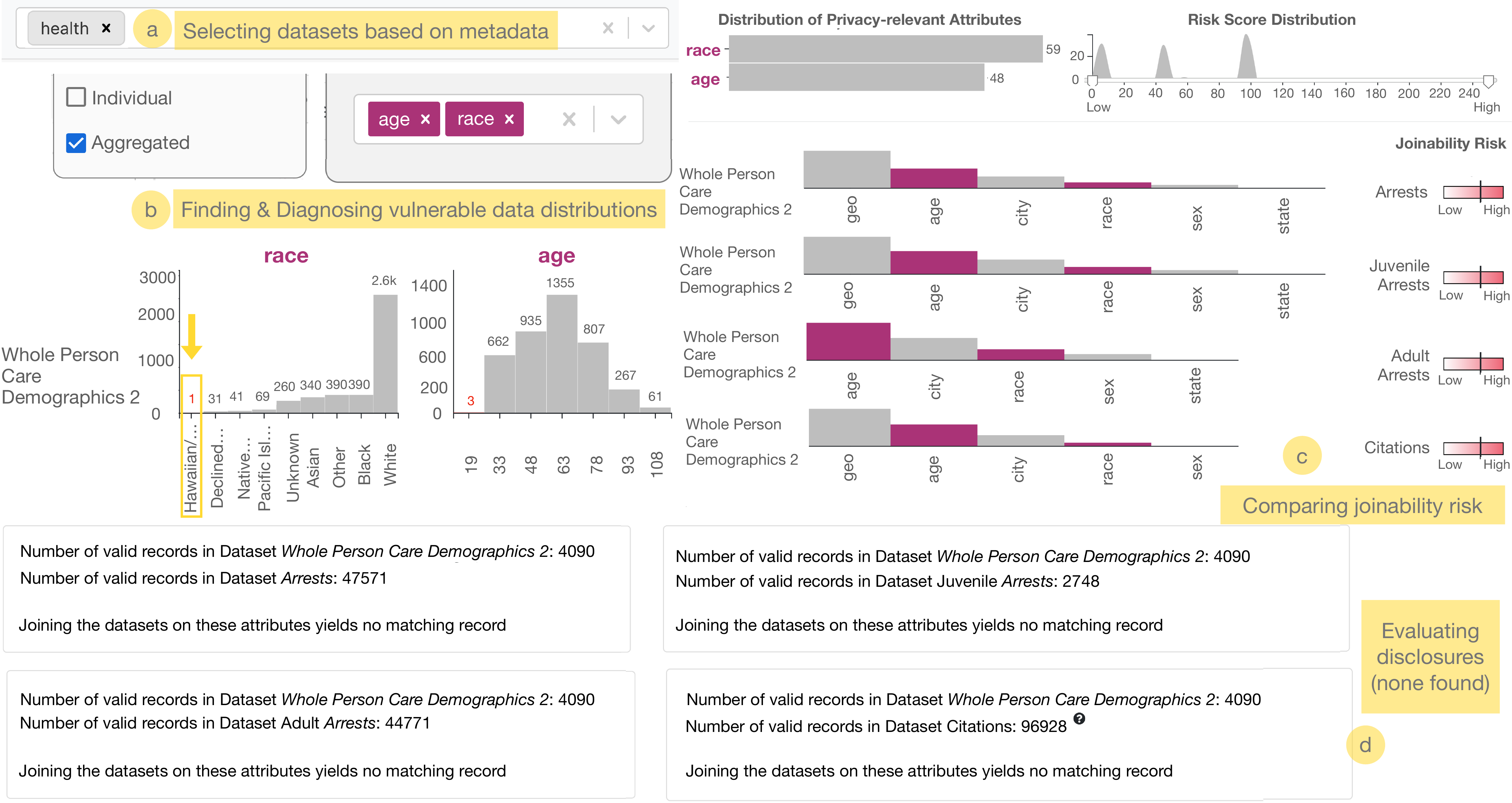}
\vspace{-7mm}
 \caption{\label{figs:case_study_vulnerability}\textbf{PRIVEE as a risk confidante for defenders:} (a) selecting datasets based on their metadata like the popular tag ``health" and their granularity of records, (b) finding and diagnosing the vulnerable data distributions and observing that there is only 1 record for the race ``Hawaiian", (c) comparing the joinability risk with the individual record-level datasets and (d) evaluating the disclosures with the top $4$ individual-level datasets and observing that there is no disclosure.}
\end{center}
\vspace{-9mm}
\end{figure*}
%The selection of the dataset pairs from the Risk Assessment View leads to the following question: can these datasets be joined to reveal sensitive information? Thus, 
The design requirement for addressing tasks T4 and T5 is to let the defenders \textit{judge the degree of sensitive information} that can ultimately be disclosed through the joins. Since an apriori definition of risky features is insufficient, PRIVEE also suggests additional features to defenders for diagnosing sensitive matches. In this section, we first discuss the methods used for evaluating the disclosures, followed by the design of the visual cues that can help evaluate them.
%these disclosures.

\subsection{Methods for disclosure evaluation}
During the red-teaming exercise, we realized that the 
%selection of the 
join key could vastly influence the disclosure of sensitive information. In this sub-section, we discuss two methods for disclosure evaluation: 
%In this sub-section, we discuss a method to evaluate the matching records between two datasets based on the selection of a particular join key and another method to update the join key based on feature suggestions from these matching records.  \\

\noindent \textbf{\textit{Based on the low frequency of matching records}}:
\textit{Matching records} are the number of records present in the joined dataset. Hence, the presence of matching records can indicate the possible disclosures at the record level. However, the number of matching records may vary according to the choice of attributes in the join key and the type of records present in the datasets. For example, when joined on attributes $x$ and $y$, dataset A and dataset B may have $200$ matching records, but when joined on the attributes $x$, $y$, and $z$, they may have only $20$ matching records. This implies that the attribute combination $x$, $y$, and $z$ have a better chance of discovering an actual disclosure than the combination $x$ and $y$. 
%PRIVEE employs an algorithm for the automatic selection of the join keys, which can also be augmented by the data defender. This algorithm selects a combination of those shared attributes which are also present in a pre-calculated list of privacy-related attributes along with the shared attributes with high entropy level.   
We have also observed that matching records may contain duplicates if the original datasets have duplicate or blank entries. 

\noindent \textbf{\textit{Based on the mutual information between the joining attributes}}:
The selection of the joining attributes is an iterative process in PRIVEE. Mutual information measures the amount of information one random variable contains about another~\cite{cover1991entropy} and quantifies the mutual dependence of the two attributes of a dataset. Hence, we use normalized mutual information to suggest other features that defenders can use for detecting disclosures. PRIVEE automatically calculates the normalized mutual information between the joining attributes and the other attributes of the joined dataset. Next, it finds the top-5 attributes with the highest mutual information score and lets defenders consider those features for detecting matches \textbf{(T5)}.

\subsection{Visual cues for evaluating disclosures}
%PRIVEE uses multiple visual cues to aid the evaluation of disclosures in the joined datasets, thus forming the Disclosure Evaluation View.
The design of the Disclosure Evaluation View follows Shneiderman's mantra~\cite{shneiderman2003eyes}, where PRIVEE first provides an overview of the matched records, then allows the defender to explore them, and finally lets them view the record details on demand. Here we discuss the 
comparative visual cues~\cite{dasgupta2020separating} that aid in disclosure evaluation:
%components of this view, which work in tandem with the other views to enable the PRIVEE workflow (Figure~\ref{figs:PRIVEE}) and help the data defender identify cases of disclosure.\\

% \begin{figure*}[t]
% \begin{center}
% \includegraphics[width=\textwidth]{figures/matches_view.pdf}
% \vspace{-4mm}
%  \caption{\label{figs:matches_view}\textbf{Disclosure Evaluation View:} While joining two datasets, data defenders can view the (a) summary of the datasets and the matching records and evaluate them using the (b) modified parallel sets visualization. The (c) rectangular boxes represent each unique category of a join key attribute, while the (d) connecting ribbons between them represent the matching records between these categories. PRIVEE also provides (e) feature suggestions that the defender can use to analyze the disclosures further.}
% \end{center}
% \vspace{-4mm}
% \end{figure*}

\noindent \textbf{\textit{Exploration of matching records}}:
Parallel Sets is a visualization method for the interactive exploration of categorical data, which shows the data frequencies instead of the individual data points~\cite{kosara2006parallel}. PRIVEE shows the matching records using a modified parallel sets visualization, as illustrated in Figure~\ref{figs:PRIVEE_Overview}d. Here, each attribute of the join key is represented using a stacked bar, where the height of the stacks represents the frequency of the different categories of that attribute. %(Figure~\ref{figs:PRIVEE_Overview}d).
In the case of a numerical attribute, a histogram replaces the stacked bar and shows its data distribution. The numerical data is then divided into four equal bins to map them with the categories of the other join key attributes. The parallel sets for the privacy-related attributes are colored in violet, while that for the other attributes are colored in grey, following the similar color scheme used in the other views. The categories across the numerical and categorical attributes are connected using ribbons. %(Figure~\ref{figs:PRIVEE_Overview}d). 
Each ribbon represents the number of records in the joined dataset belonging to both categories. A simple click interaction on any of these ribbons opens a pop-up window showing the details of the records represented by the selected line.

This design helps detect both identity and attribute disclosures through the matching records \textbf{(T4)}. The thickness of the line may represent the identity disclosure, while the height of the stacked bar shows the attribute disclosure. For example, if there is only one record with a certain combination of all the join key attributes, this would be represented by a thin ribbon across the parallel sets visualization. This may potentially lead to identity disclosure if an individual is uniquely identified with this combination of the join key. Suppose if an attribute has only one category, then the corresponding stack height would cover all the height allocated to a certain attribute, revealing that all the individuals belonging to both the datasets have a particular feature and leading to attribute disclosure. This Disclosure Evaluation View helps the data defenders ascertain the degree of the sensitive information disclosed by visualizing the overall relationship between the different attributes of the matching records yet retaining the granularity of the dataset at the record level.

\noindent \textbf{\textit{Suggesting potential joining attributes}}:
PRIVEE uses bar charts and histograms to encode the top-5 features with high mutual information with the join key attributes. These suggestions are positioned on the left and right-hand sides of the parallel sets, representing the feature suggestions from either of the datasets~(Figure~\ref{figs:PRIVEE_Overview}d). The privacy-related attributes are also highlighted in violet, while the others are colored in grey, following a color scheme similar to the interface's other views. Selecting any attributes from the feature suggestions would also update this visualization to include the newly selected attributes. 
These attributes can be used as suggestions for improving the initial set of joining attributes \textbf{(T5)}. The data distributions and the ranking of the attributes help boost defenders' understanding of the risky feature set that can be used as the join key.

%\input{sections/data_collection_and_analysis_process}
%\input{sections/design_of_privee}
% \section{Evaluation}\label{section:evaluation}
% In this section, we describe two case studies with our collaborators which demonstrate the effectiveness of PRIVEE in analyzing the open data ecosystem for disclosure risks. %One of the collaborators observed that PRIVEE could help data defenders triage the different combinations of open datasets and eventually find disclosure risks, while the other collaborator pointed out that PRIVEE can also act as a risk confidante for the data defenders.

\section{Case Study: Risk confidante} \label{section:case_study_vulnerability}

\begin{figure*}
\begin{center}
\includegraphics[width=\textwidth]{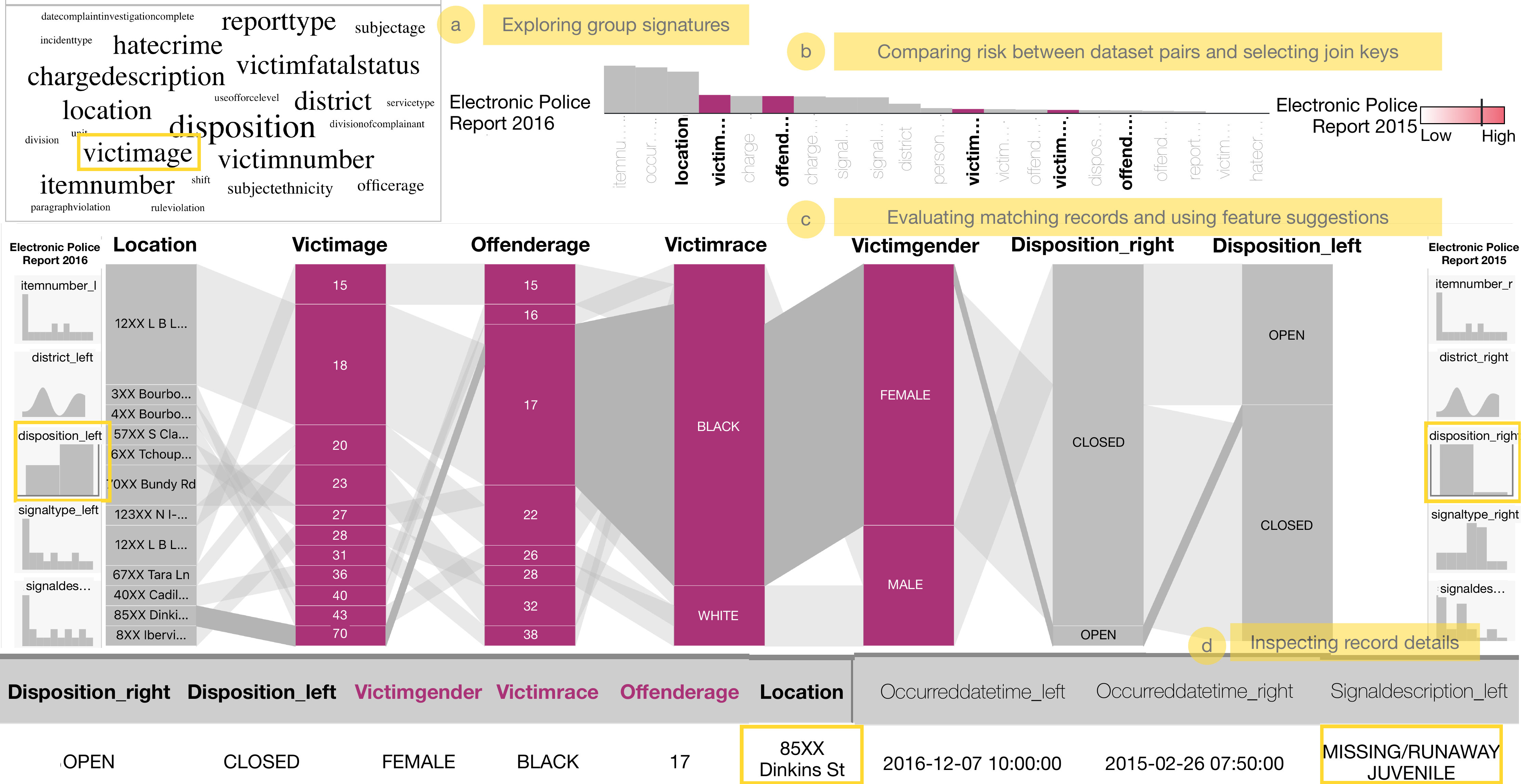}
\vspace{-6mm}
 \caption{\label{figs:case_study2}\textbf{PRIVEE as a trusted informer for defenders:} (a) understanding group signatures and updating privacy-related attributes, (b)comparing the risk between dataset pairs, (c) evaluating the matching records using the feature suggestions shows that only one incident was open in 2015 but closed in 2016, (d) inspecting record details shows that a runaway juvenile can be identified despite the location being partially masked.}
\end{center}
\vspace{-9mm}
\end{figure*}

We report a case study that our data privacy collaborator and co-author co-developed using the web interface of PRIVEE. He is a senior researcher with more than $15$ years of experience in privacy-preserving data analysis and used PRIVEE as a privacy auditor. Specifically, he wanted to determine if there are any disclosure risks with the health-related datasets published in the open data portals and validate the role of PRIVEE as a risk confidante for data defenders.

% Our collaborator knew that the dataThe officials of various Open Government Data (OGD)~\cite{OpenGovernmentData:online} initiatives assured our collaborator that the citizens' data are not published directly to its open data portals~\cite{SMCDatahub:online, CountyofSantaClara:online, SantaMonica:online, RiversideCounty:online}. Instead, they are aggregated using different attributes like age, race, sex, etc. 
%He was worried if his or his family's health data could be disclosed online, so he selected the aggregated datasets tagged with the keyword ``health" in the interface PRIVEE and then selected the data portal of the San Mateo county (see Figure \ref{figs:case_study_vulnerability}a).
Our collaborator selected the aggregated datasets in the interface PRIVEE along with the privacy-related attributes \textit{age} and \textit{race}; and then filtered them with the keyword ``health" (see Figure \ref{figs:case_study_vulnerability}a).
%He also selected few privacy-related attributes like \textit{age} and \textit{race} since he was informed that the datasets are aggregated using these attributes (Figure \ref{figs:case_study_vulnerability}b). % (\textbf{T1}). 
He also enabled the Vulnerable Datasets switch to check if there are any vulnerabilities in the data distributions of these datasets. At this point, our collaborator observed that the first few clusters do not have such vulnerable datasets. However, the fourth cluster has the dataset \textit{Whole Person Care Demographics 2} \cite{WholePersonDemographics2:online} from the open data portal of San Mateo county~\cite{SMCDatahub:online}. This dataset had only $1$ record where the race was Hawaiian~(Figure \ref{figs:case_study_vulnerability}c) (\textbf{T2}).
% Since there were only a few datasets in this selection, he observed that there is only $1$ dataset with the attribute \textit{race}. Hence, he enabled the Vulnerable Datasets switch to check if there are any vulnerabilities in these datasets. 
This was a significant cause of concern
%for our collaborator
since if somebody knows a person in that county who identified as Hawaiian, then any dataset with a similar race category could potentially expose her health records. Thus, he started analyzing the risk of joining this dataset with all the individual-record level datasets available through PRIVEE, as shown in Figure \ref{figs:case_study_vulnerability}c (\textbf{T3}). 
%Since there would be around 151 dataset combinations (the number of individual-record level datasets available), he decided to filter those combinations with a higher risk score ($\geq 130$), using the filter available in the risk score distribution histogram.
He decided to join these dataset pairs on the selected privacy-related attributes and the location attribute \textit{geocodedcolumnn} since he wished to find datasets containing information relevant to this location. He observed that none of the top-4 dataset pairs yield any matching record when joined on these attributes (Figure \ref{figs:case_study_vulnerability}d). Thus, our collaborator concluded that though this aggregated dataset has a meager count of a particular race, it does not lead to any disclosure (\textbf{T4}). He also analyzed a few other vulnerable datasets similarly but found no disclosures. Thus, PRIVEE acts as a risk confidante for the data defenders where they can analyze the disclosure risks for the vulnerable datasets in the presence of other open datasets. He also observed that he had not seen a tool with similar capabilities for interactive risk calibration and triage and commented: ``\textit{this is a great visual tool to explore privacy risks of open data, with the ability to visualize privacy risk across datasets in a dynamic manner}".

\section{Case Study: Trusted informer}\label{section:case_study_id_disclosure}

We report a case study that a researcher developed using the PRIVEE web interface. He is a senior researcher and university professor with over 25 years of experience in the fields of big data, cyber security, and scientific visualization.
%Though he was consulted during the ideation stage of the PRIVEE workflow, he was presented with the interface for the first time during the case study process. 
%He focused on finding examples of disclosures to validate the role of PRIVEE as a trusted informer for the data defenders.
He focused on validating the role of PRIVEE as a trusted informer for the data defenders.
%through PRIVEE and subsequently audit their data repository. 

The researcher started by choosing the New Orleans Open Data portal~\cite{NewOrleans:online} and observed $7$ datasets on the Projection View, which were so similar in their attribute space that they were displayed using an overlapping circle with the number of datasets inscribed. Using the attribute distribution bar chart, he observed that none of the default privacy-related attributes (age, race, gender) were present in this group of datasets. However, on analyzing the word cloud, he made an interesting observation that attributes like \textit{victim age} and \textit{offender age} were present in these datasets, as shown in Figure~\ref{figs:case_study2}a~\textbf{(T1)}. Since, from his background knowledge, he knew that these attributes are generally present in police datasets, he updated the list of privacy-related attributes to select some of the similar attributes like \textit{victim age}, \textit{victim gender}, \textit{victim race}, and \textit{offender age}. As PRIVEE helps to triage the joinable groups of datasets based on the data defender's definition of privacy relevance, the Projection View was updated to reflect the change in privacy-related attributes. 

He selected all these seven datasets in order to compare the joinability risks of the $21$ possible pairwise combinations in the Risk Assessment View~\textbf{(T3)}. Since he wanted to focus only on the high-risk pairs, he filtered out the low-risk pairs using the Risk Score Distribution histogram. 
%Next, he selected the first pair of datasets, \textit{Electronic Police Report 2018} and \textit{Electronic Police Report 2019}, and observed that some attributes like \textit{location}, \textit{victim age}, \textit{offender age}, \textit{victim race}, and \textit{offender race} were automatically suggested as the attributes for the join key. 
%PRIVEE suggests these attributes based on a custom algorithm that focuses on the attributes profile and the definition of privacy-relevance by the data defender (Section \ref{sub_section:metrics_joinability_risk}). 
Joining the first pair of datasets, the researcher observed that there are no matching records between them.% these datasets.

%He further selected the next pair of datasets, namely \textit{Electronic Police Report 2018} and \textit{Electronic Police Report 2016}, and joined them using the join key attributes suggested by PRIVEE. He observed that these datasets have $23$ matching records, which can be evaluated using the modified parallel sets visualization in the Disclosure Evaluation View \textbf{(T4)}. He made an interesting observation that matching records have a similar racial profile between the victim and the offender. 
Next, he selected a pair of datasets, namely \textit{Electronic Police Report $2016$} and \textit{Electronic Police Report $2015$}, but augmented the PRIVEE-suggested join key attributes and made the following selection: \textit{location}, \textit{victim age}, \textit{offender age}, \textit{victim race}, \textit{victim gender}, \textit{offender gender}, as illustrated in Figure~\ref{figs:case_study2}b~\textbf{(T3)}. He joined these datasets and observed $14$ matching records in the Disclosure Evaluation View. He inspected further details about a certain record and observed that a $22$-year-old black male was charged with attempted robbery with a gun against a $27$-year-old white male at $6XX$ Tchoupitoulas St on $13$\textsuperscript{th} July $2015$ at $01:00$ hrs and again on $30$\textsuperscript{th} April $2016$ at $03:00$ hrs with attempted simple robbery~\textbf{(T4)}. Next, from the feature suggestions offered by PRIVEE~\textbf{(T5)}, he selected the attribute \textit{disposition}, which shows the status of a particular incident. He observed that only one record was open in $2015$ but closed in $2016$~(Figure~\ref{figs:case_study2}c). On inspecting further details, as shown in Figure~\ref{figs:case_study2}d, he found out that an incident of a runaway female juvenile of age $17$ was reported at $85XX$ Dinkins St on $26$\textsuperscript{th} February $2015$, and the same incident was closed through a supplemental report one and half years later on $7$\textsuperscript{th} December $2016$~\textbf{(T4)}.

The researcher concluded that this is an example of identity disclosure where individuals were identified using PRIVEE even when the addresses were partially masked in de-identified datasets. He was also shown an earlier version of the PRIVEE interface during the case study. He commented that the new changes ``\textit{improved the rich functionalities}" of PRIVEE and added that this interface could ``\textit{help experienced data custodians analyze disclosure risks and potentially find examples of disclosures}".

\section{Discussion}
% Impact (reinforce first two)
% * First study demonstrating open data vulnerability
% * Plug into a data stream, trusted informer and risk profiler- during dataset release or after release, when the links change 
% * Possible to use more automation with ML learning about privacy-relevance and feature use and automatically re-weighting for generating  the projection plots

% Ethics
% * Like all other technical contributions on modeling adversarial attacks, can be used in a malicious way
% * Limited release of the software after deliberate with our collaborators

% Automation augmentation
% * Pairs to non-pairs
% * Seed set - reasonable assumption as we focus on the most vulnerable datasets, it is can open challenge what privacy relevance means in an open data context and how to automate that process.
% * Other reports?

% Limitation and Future Work
% * scalability
% * pre-fetching

%The two case studies demonstrate the success of PRIVEE in helping defenders discover disclosure risks in real-world open data ecosystems. 
When plugged into the open data stream, PRIVEE can act as both a risk profiler and a trusted informer that oversees risks while providing an appropriate level of control to defenders for integrating their domain knowledge using an end-to-end workflow. One of the lessons learned during this design study is that an interface helping defenders evaluate disclosures should enable seamless communication across sources and implications of risks while responding to the myriad definitions of privacy relevance. PRIVEE is bootstrapped by a default view that quickly adapts to the data defenders' inputs, allowing them to leverage appropriate levels of control while automating parts of the analysis process.

In its current implementation, one of the limitations of PRIVEE is scalability, concerning the number of records of each processed dataset and the size of the seed input that is used for bootstrapping. We have limited the number of records to $100,000$ to avoid interaction latency. 
%While currently, we are processing more than $400$ datasets, our experiments have shown that PRIVEE can efficiently handle the computation for $1000$ datasets.
%the computation required to process many join combinations of $1000$ datasets. 

There is also the need to incorporate greater automation in the selection of privacy-relevant, personal datasets without manual intervention. During this design study process, we learned that automation of this workflow is inherently challenging as privacy-relevance is subjective and open data are noisy; hence, training a model to mimic human judgment is difficult. Our approach of specifying a seed set outside the PRIVEE workflow is an important methodological choice allowing us to focus on the most vulnerable datasets and anticipated attack scenarios. Currently, PRIVEE only assesses joinability risk between pairs of datasets. It is certainly possible that there could be other scenarios like when multiple datasets are joined progressively, the risks propagate through the links.
%There could also be other more complex attack scenarios.
However, based on the feedback of our data privacy collaborator, we consider the risk scenarios handled in PRIVEE to be the necessary first steps toward assessing more complex combinations and variants of disclosure risks. 

\section{Conclusion}
PRIVEE, the visual risk inspection workflow described in this design study paper, is a first step towards allowing data defenders both the control and efficiency needed to minimize disclosure risks from the joinability of open datasets. Through our case studies with data privacy experts, we demonstrated a key takeaway that the visualizations and interactions were effective in end-to-end exploration and diagnosis of actual disclosure of sensitive information or identity of individuals. As an ongoing and future work, we will be exploring disclosure risks beyond joinable pairs. 
We will further augment our workflow with intelligent and scalable data processing capabilities in collaboration with big data experts. 
We also plan to conduct controlled studies for evaluating the usability of PRIVEE and its components with real-world cyber defenders.  

\acknowledgments{
The work reported in this publication was supported by the National Science Foundation~(CNS-2027789) and the National Institutes of Health~(R35GM134927). The content is solely the responsibility of the authors and does not necessarily represent the official views of the agencies funding the research.}

\bibliographystyle{abbrv-doi}
\balance
\bibliography{bib}

\begin{thebibliography}{10}

\bibitem{alfalayleh2014quantifying}
M.~Alfalayleh and L.~Brankovic.
\newblock Quantifying privacy: A novel entropy-based measure of disclosure
  risk.
\newblock In {\em International Workshop on Combinatorial Algorithms}, pp.
  24--36. Springer, 2014.

\bibitem{almeida2019word}
F.~Almeida and G.~Xex{\'e}o.
\newblock Word embeddings: A survey.
\newblock {\em arXiv preprint arXiv:1901.09069}, 2019.

\bibitem{ankerst1999optics}
M.~Ankerst, M.~M. Breunig, H.-P. Kriegel, and J.~Sander.
\newblock Optics: Ordering points to identify the clustering structure.
\newblock {\em ACM Sigmod record}, 28(2):49--60, 1999.

\bibitem{bakarov2018survey}
A.~Bakarov.
\newblock A survey of word embeddings evaluation methods.
\newblock {\em arXiv preprint arXiv:1801.09536}, 2018.

\bibitem{bhattacharjee2020privacy}
K.~Bhattacharjee, M.~Chen, and A.~Dasgupta.
\newblock Privacy-preserving data visualization: Reflections on the state of
  the art and research opportunities.
\newblock In {\em Computer Graphics Forum}, vol.~39, pp. 675--692. Wiley Online
  Library, Norrköping, Sweden, 2020.

\bibitem{PRIVEE_NJIT_Dataset}
K.~Bhattacharjee, A.~Islam, J.~Vaidya, and A.~Dasgupta.
\newblock {PRIVEE-NJIT dataset}.
\newblock \url{https://doi.org/10.7910/DVN/VHOR3V}, 2022. doi: {{%
10\hspace{.1pt}\discretionary{.}{%
}{.}\hspace{.4pt}7910\discretionary{/}{%
}{/}DVN\discretionary{/}{%
}{/}VHOR3V}}


\bibitem{calinski1974dendrite}
T.~Cali{\'n}ski and J.~Harabasz.
\newblock A dendrite method for cluster analysis.
\newblock {\em Communications in Statistics-theory and Methods}, 3(1):1--27,
  1974.

\bibitem{Ourhisto94:online}
O.~D. Charter.
\newblock Our history - international open data charter.
\newblock \url{https://opendatacharter.net/our-history/}, 2020.
\newblock (Accessed on 07/19/2021).

\bibitem{chia2019khyperloglog}
P.~H. Chia, D.~Desfontaines, I.~M. Perera, D.~Simmons-Marengo, C.~Li, W.-Y.
  Day, Q.~Wang, and M.~Guevara.
\newblock Khyperloglog: Estimating reidentifiability and joinability of large
  data at scale.
\newblock In {\em 2019 IEEE Symposium on Security and Privacy (SP)}, pp.
  350--364. IEEE, 2019.

\bibitem{DallasOpenData:online}
{City of Dallas Open Data}.
\newblock \url{https://www.dallasopendata.com/}.
\newblock (Accessed on 10/05/2021).

\bibitem{cover1999elements}
T.~M. Cover.
\newblock {\em Elements of information theory}.
\newblock John Wiley \& Sons, 1999.

\bibitem{cover1991entropy}
T.~M. Cover, J.~A. Thomas, et~al.
\newblock Entropy, relative entropy and mutual information.
\newblock {\em Elements of information theory}, 2(1):12--13, 1991.

\bibitem{culnane2017health}
C.~Culnane, B.~I. Rubinstein, and V.~Teague.
\newblock Health data in an open world.
\newblock {\em arXiv preprint arXiv:1712.05627}, 2017.

\bibitem{dai2015document}
A.~M. Dai, C.~Olah, and Q.~V. Le.
\newblock Document embedding with paragraph vectors.
\newblock {\em arXiv preprint arXiv:1507.07998}, 2015.

\bibitem{dasgupta2019}
A.~Dasgupta, R.~Kosara, and M.~Chen.
\newblock Guess me if you can: A visual uncertainty model for transparent
  evaluation of disclosure risks in privacy-preserving data visualization.
\newblock {\em VizSec}, pp. 1--10, 2019.

\bibitem{dasgupta2014opportunities}
A.~Dasgupta, E.~Maguire, A.-R. Alfie, and M.~Chen.
\newblock Opportunities and challenges for privacy-preserving visualization of
  electronic health record data.
\newblock In {\em Proceedings of IEEE VIS 2014 Workshop on Visualization of
  Electronic Health Records}, 2014.

\bibitem{dasgupta2020separating}
A.~Dasgupta, H.~Wang, O.~Nancy, and S.~Burrows.
\newblock Separating the wheat from the chaff: Comparative visual cues for
  transparent diagnostics of competing models.
\newblock {\em IEEE Transactions on Visualization and Computer Graphics}, 2020.

\bibitem{davies1979cluster}
D.~L. Davies and D.~W. Bouldin.
\newblock A cluster separation measure.
\newblock {\em IEEE transactions on pattern analysis and machine intelligence},
  pp. 224--227, 1979.

\bibitem{de2013unique}
Y.-A. De~Montjoye, C.~A. Hidalgo, M.~Verleysen, and V.~D. Blondel.
\newblock Unique in the crowd: The privacy bounds of human mobility.
\newblock {\em Scientific reports}, 3(1):1--5, 2013.

\bibitem{de2015unique}
Y.-A. De~Montjoye, L.~Radaelli, V.~K. Singh, and A.~S. Pentland.
\newblock Unique in the shopping mall: On the reidentifiability of credit card
  metadata.
\newblock {\em Science}, 347(6221):536--539, 2015.

\bibitem{diaz2002towards}
C.~Diaz, S.~Seys, J.~Claessens, and B.~Preneel.
\newblock Towards measuring anonymity.
\newblock In {\em International Workshop on Privacy Enhancing Technologies},
  pp. 54--68. Springer, 2002.

\bibitem{dong2021efficient}
Y.~Dong, K.~Takeoka, C.~Xiao, and M.~Oyamada.
\newblock Efficient joinable table discovery in data lakes: A high-dimensional
  similarity-based approach.
\newblock In {\em 2021 IEEE 37th International Conference on Data Engineering
  (ICDE)}, pp. 456--467. IEEE, 2021.

\bibitem{douriez2016anonymizing}
M.~Douriez, H.~Doraiswamy, J.~Freire, and C.~T. Silva.
\newblock Anonymizing nyc taxi data: Does it matter?
\newblock In {\em 2016 IEEE international conference on data science and
  advanced analytics (DSAA)}, pp. 140--148. IEEE, 2016.

\bibitem{dwork2019differential}
C.~Dwork.
\newblock Differential privacy and the us census.
\newblock In {\em Proceedings of the 38th ACM SIGMOD-SIGACT-SIGAI Symposium on
  Principles of Database Systems}, pp. 1--1, 2019.

\bibitem{dwork2014algorithmic}
C.~Dwork, A.~Roth, et~al.
\newblock The algorithmic foundations of differential privacy.
\newblock {\em Found. Trends Theor. Comput. Sci.}, 9(3-4):211--407, 2014.

\bibitem{eick2002visual}
S.~G. Eick and A.~F. Karr.
\newblock Visual scalability.
\newblock {\em Journal of Computational and Graphical Statistics},
  11(1):22--43, 2002.

\bibitem{ester1996density}
M.~Ester, H.-P. Kriegel, J.~Sander, X.~Xu, et~al.
\newblock A density-based algorithm for discovering clusters in large spatial
  databases with noise.
\newblock In {\em kdd}, vol. 96,34, pp. 226--231, 1996.

\bibitem{FortLauderdaleOpenData:online}
{City of Fort Lauderdale Police Department Open Data}.
\newblock \url{https://fortlauderdale.data.socrata.com/}.
\newblock (Accessed on 10/05/2021).

\bibitem{fung2010privacy}
B.~C. Fung, K.~Wang, R.~Chen, and P.~S. Yu.
\newblock Privacy-preserving data publishing: A survey of recent developments.
\newblock {\em ACM Computing Surveys (Csur)}, 42(4):1--53, 2010.

\bibitem{green2017open}
B.~Green, G.~Cunningham, A.~Ekblaw, P.~Kominers, A.~Linzer, and S.~P. Crawford.
\newblock Open data privacy.
\newblock {\em Berkman Klein Center Research Publication}, pp. 17--07, 2017.

\bibitem{opendatadefinition}
O.~K. O.~D. Group.
\newblock Open definition, 2015.

\bibitem{hutchins2011intelligence}
E.~M. Hutchins, M.~J. Cloppert, R.~M. Amin, et~al.
\newblock Intelligence-driven computer network defense informed by analysis of
  adversary campaigns and intrusion kill chains.
\newblock {\em Leading Issues in Information Warfare \& Security Research},
  1(1):80, 2011.

\bibitem{kao2017using}
C.-H. Kao, C.-H. Hsieh, Y.-F. Chu, Y.-T. Kuang, and C.-K. Yang.
\newblock Using data visualization technique to detect sensitive information
  re-identification problem of real open dataset.
\newblock {\em Journal of Systems Architecture}, 80:85--91, 2017.

\bibitem{kosara2006parallel}
R.~Kosara, F.~Bendix, and H.~Hauser.
\newblock Parallel sets: Interactive exploration and visual analysis of
  categorical data.
\newblock {\em IEEE transactions on visualization and computer graphics},
  12(4):558--568, 2006.

\bibitem{kum2019enhancing}
H.-C. Kum, E.~D. Ragan, G.~Ilangovan, M.~Ramezani, Q.~Li, and C.~Schmit.
\newblock Enhancing privacy through an interactive on-demand incremental
  information disclosure interface: Applying $\{$Privacy-by-Design$\}$ to
  record linkage.
\newblock In {\em Fifteenth Symposium on Usable Privacy and Security (SOUPS
  2019)}, pp. 175--189, 2019.

\bibitem{lavrenovs2016privacy}
A.~Lavrenovs and K.~Podins.
\newblock Privacy violations in riga open data public transport system.
\newblock In {\em 2016 IEEE 4th Workshop on Advances in Information, Electronic
  and Electrical Engineering (AIEEE)}, pp. 1--6. IEEE, 2016.

\bibitem{lloyd2011human}
D.~Lloyd and J.~Dykes.
\newblock Human-centered approaches in geovisualization design: Investigating
  multiple methods through a long-term case study.
\newblock {\em IEEE transactions on visualization and computer graphics},
  17(12):2498--2507, 2011.

\bibitem{miller2018making}
R.~J. Miller, F.~Nargesian, E.~Zhu, C.~Christodoulakis, K.~Q. Pu, and
  P.~Andritsos.
\newblock Making open data transparent: Data discovery on open data.
\newblock {\em IEEE Data Eng. Bull.}, 41(2):59--70, 2018.

\bibitem{NewOrleans:online}
City of new orleans | open data.
\newblock \url{https://datadriven.nola.gov/home/}.
\newblock (Accessed on 11/02/2021).

\bibitem{nouwens2020dark}
M.~Nouwens, I.~Liccardi, M.~Veale, D.~Karger, and L.~Kagal.
\newblock Dark patterns after the gdpr: Scraping consent pop-ups and
  demonstrating their influence.
\newblock In {\em Proceedings of the 2020 CHI conference on human factors in
  computing systems}, pp. 1--13, 2020.

\bibitem{NYCOpenData:online}
{NYC Open Data}.
\newblock \url{https://opendata.cityofnewyork.us/}.
\newblock (Accessed on 10/05/2021).

\bibitem{oganian2003posteriori}
A.~Oganian and J.~Domingo~Ferrer.
\newblock A posteriori disclosure risk measure for tabular data based on
  conditional entropy.
\newblock {\em SORT. 2003, Vol. 27, N{\'u}m. 2 [July-December]}, 2003.

\bibitem{ohm2009broken}
P.~Ohm.
\newblock Broken promises of privacy: Responding to the surprising failure of
  anonymization.
\newblock {\em Ucla L. Rev.}, 57:1701, 2009.

\bibitem{KansasCityOpenData:online}
{Open Data Kansas City}.
\newblock \url{https://data.kcmo.org/}.
\newblock (Accessed on 10/05/2021).

\bibitem{rocher2019estimating}
L.~Rocher, J.~M. Hendrickx, and Y.-A. De~Montjoye.
\newblock Estimating the success of re-identifications in incomplete datasets
  using generative models.
\newblock {\em Nature communications}, 10(1):1--9, 2019.

\bibitem{rousseeuw1987silhouettes}
P.~J. Rousseeuw.
\newblock Silhouettes: a graphical aid to the interpretation and validation of
  cluster analysis.
\newblock {\em Journal of computational and applied mathematics}, 20:53--65,
  1987.

\bibitem{rubinstein2016anonymization}
I.~S. Rubinstein and W.~Hartzog.
\newblock Anonymization and risk.
\newblock {\em Wash. L. Rev.}, 91:703, 2016.

\bibitem{ruggles2019differential}
S.~Ruggles, C.~Fitch, D.~Magnuson, and J.~Schroeder.
\newblock Differential privacy and census data: Implications for social and
  economic research.
\newblock In {\em AEA papers and proceedings}, vol. 109, pp. 403--08, 2019.

\bibitem{schubert2018improving}
E.~Schubert and M.~Gertz.
\newblock Improving the cluster structure extracted from optics plots.
\newblock In {\em LWDA}, 2018.

\bibitem{schubert2017dbscan}
E.~Schubert, J.~Sander, M.~Ester, H.~P. Kriegel, and X.~Xu.
\newblock Dbscan revisited, revisited: why and how you should (still) use
  dbscan.
\newblock {\em ACM Transactions on Database Systems (TODS)}, 42(3):1--21, 2017.

\bibitem{sekara2021temporal}
V.~Sekara, L.~Alessandretti, E.~Mones, and H.~Jonsson.
\newblock Temporal and cultural limits of privacy in smartphone app usage.
\newblock {\em Scientific reports}, 11(1):1--9, 2021.

\bibitem{serjantov2002towards}
A.~Serjantov and G.~Danezis.
\newblock Towards an information theoretic metric for anonymity.
\newblock In {\em International Workshop on Privacy Enhancing Technologies},
  pp. 41--53. Springer, 2002.

\bibitem{shneiderman2003eyes}
B.~Shneiderman.
\newblock The eyes have it: A task by data type taxonomy for information
  visualizations.
\newblock In {\em The craft of information visualization}, pp. 364--371.
  Elsevier, 2003.

\bibitem{SMCDatahub:online}
{SMC Datahub}.
\newblock \url{https://datahub.smcgov.org/}.
\newblock (Accessed on 10/07/2021).

\bibitem{Spacy}
spacy · industrial-strength natural language processing in python.
\newblock \url{https://spacy.io/}.
\newblock (Accessed on 11/14/2021).

\bibitem{thongtan2019sentiment}
T.~Thongtan and T.~Phienthrakul.
\newblock Sentiment classification using document embeddings trained with
  cosine similarity.
\newblock In {\em Proceedings of the 57th Annual Meeting of the Association for
  Computational Linguistics: Student Research Workshop}, pp. 407--414, 2019.

\bibitem{van2008visualizing}
L.~Van~der Maaten and G.~Hinton.
\newblock Visualizing data using t-sne.
\newblock {\em Journal of machine learning research}, 9(11), 2008.

\bibitem{vassilvitskii2006k}
S.~Vassilvitskii and D.~Arthur.
\newblock k-means++: The advantages of careful seeding.
\newblock In {\em Proceedings of the eighteenth annual ACM-SIAM symposium on
  Discrete algorithms}, pp. 1027--1035, 2006.

\bibitem{GoogleDLP:online}
Visualizing re-identification risk using data studio | data loss prevention
  documentation | google cloud.
\newblock \url{https://cloud.google.com/dlp/docs/visualizing_re-id_risk}.
\newblock (Accessed on 06/28/2022).

\bibitem{WholePersonDemographics2:online}
{Whole Person Care Demographics 2 | SMC Datahub}.
\newblock
  \url{https://datahub.smcgov.org/dataset/Whole-Person-Care-Demographics-2/qqdq-93h5}.
\newblock (Accessed on 10/07/2021).

\bibitem{wilkinson2016fair}
M.~D. Wilkinson, M.~Dumontier, I.~J. Aalbersberg, G.~Appleton, M.~Axton,
  A.~Baak, N.~Blomberg, J.-W. Boiten, L.~B. da~Silva~Santos, P.~E. Bourne,
  et~al.
\newblock The fair guiding principles for scientific data management and
  stewardship.
\newblock {\em Scientific data}, 3(1):1--9, 2016.

\bibitem{zang2011anonymization}
H.~Zang and J.~Bolot.
\newblock Anonymization of location data does not work: A large-scale
  measurement study.
\newblock In {\em Proceedings of the 17th annual international conference on
  Mobile computing and networking}, pp. 145--156, 2011.

\bibitem{zenko2015red}
M.~Zenko.
\newblock {\em Red Team: How to succeed by thinking like the enemy}.
\newblock Basic Books, 2015.

\bibitem{zhang1996birch}
T.~Zhang, R.~Ramakrishnan, and M.~Livny.
\newblock Birch: an efficient data clustering method for very large databases.
\newblock {\em ACM sigmod record}, 25(2):103--114, 1996.

\end{thebibliography}
\end{document}